\documentclass[preprint,12pt,authoryear]{elsarticle}

\usepackage{amssymb}
\usepackage{xcolor}
\usepackage{tablefootnote}
\usepackage{amsmath}

\usepackage{lineno}

\journal{Icarus}

\newcommand{\altwosix}{\mbox{${}^{26}{\rm Al}$}}
\newcommand{\mnfivethree}{\mbox{${}^{53}{\rm Mn}$}}

\newcommand{\alratio}{\mbox{${}^{26}{\rm Al}/{}^{27}{\rm Al}$}}

\newcommand{\mnratio}{\mbox{${}^{53}{\rm Mn}/{}^{55}{\rm Mn}$}}
\newcommand{\hfratio}{\mbox{${}^{182}{\rm Hf}/{}^{180}{\rm Hf}$}}

\newcommand{\uratio}{\mbox{${}^{238}{\rm U}/{}^{235}{\rm U}$}} 
\newcommand{\chisq}{\mbox{$\chi_{\nu}^{2}$}}
\newcommand{\permil}{\mbox{\text{\textperthousand}}}
\newenvironment{Steve}{\color{black}}{\ignorespacesafterend}

\newenvironment{DRD}{\color{black}}{\ignorespacesafterend}
\newenvironment{EDIT}{\color{black}}{\ignorespacesafterend}
\graphicspath{{./}{figures/}}

\begin{document}

\begin{frontmatter}



\title{Statistical Chronometry of Meteorites. I. A Test of ${}^{26}{\rm Al}$ homogeneity and the Pb-Pb age of the Solar System's $t\!\!=\!\!0$.}

\author[inst1]{Steven J. Desch}
% ORC-ID  0000-0002-1571-0836
%\correspondingauthor{Steve Desch}
%\email{steve.desch@asu.edu}

\affiliation[inst1]{organization={School of Earth and Space Exploration, Arizona State University},%Department and Organization
            addressline={PO Box 871404}, 
            city={Tempe},
            postcode={85287-1404}, 
            state={Arizona},
            country={USA}}

\author[inst2]{Daniel R. Dunlap}
\affiliation[inst2]{organization={Oak Ridge National Laboratory}, 
            addressline={1 Bethel Valley Rd}, 
            city={Oak Ridge},
            postcode={37830}, 
            state={Tennessee},
            country={USA}}

\author[inst6]{Emilie T. Dunham}
% ORC-ID 0000-0002-0786-7307

\affiliation[inst6]{organization={Department of Earth, Planetary and Space Sciences, University of California, Los Angeles},
            addressline={PO Box 951567},
            city={Los Angeles},
            postcode={90095-1567},
            state={California},
            country={USA}}
            
\author[inst3]{Curtis D. Williams}

\affiliation[inst3]{organization={Earth and Planetary Sciences Department, University of California, Davis},%Department and Organization
            addressline={One Shields Ave.}, 
            city={Davis},
            postcode={95616}, 
            state={California},
            country={USA}}
     
\author[inst4]{Prajkta Mane}
% ORC-ID 0000-0002-6918-2653

\affiliation[inst4]{organization={Lunar and Planetary Institute, USRA},
            addressline={3600 Bay Area Blvd.},
            city={Houston},
            postcode={77058},
            state={Texas},
            country={USA}}
%\affiliation[inst5]{organization={Astromaterials Research Exploration Sciences, NASA Johnson Space Center},
%            addressline={2101 NASA Parkway},
%            city={Houston},
%            postcode={77058},
%            state={Texas},
%            country={USA}}

\begin{abstract}
{\Steve
We use rapidly cooled achondrites to test the assumption of $\altwosix$ homogeneity in the solar nebula, by checking if there is a single value of $t_{\rm SS}$, the absolute ``Pb-Pb" age of the Solar System's $t\!\!=\!\!0$, that makes concordant their ages from the Al-Mg and Pb-Pb systems.
We find that values $t_{\rm SS} = 4568.42 \pm 0.24$ Myr do make these ages concordant, and therefore the hypothesis of homogeneous $\altwosix$ is not falsified. 
This age, defined to be when the solar nebula had $(\alratio) = 5.23 \times 10^{-5}$, is significantly older than the $\approx$ 4567.3 Myr inferred from direct measurements of Pb-Pb ages in CAIs.
Discrepancies between the Al-Mg and Pb-Pb chronometers in chondrules and CAIs have previously been interpreted as arising from heterogeneities in $\altwosix$, under the presumption that the Al-Mg and Pb-Pb systems in CAIs closed simultaneously.
We examine this assumption and show that resetting is to be expected in CAIs. In particular, we quantitatively demonstrate that it is plausible that Pb-Pb ages of CAIs were reset at late times, without resetting the earlier Al-Mg ages, if they were transiently heated in the same manner as chondrules. 
%We critically examine the assumption of simultaneous closure in two other ``carbonaceous" achondrites. 
We critically examine Pb-Pb isochrons, refining data and suggesting best practices for their calculation and reporting.
We  advocate reporting chronometry as times of formation after $t\!\!=\!\!0$ rather than absolute ages, as only the former is useful for astrophysical models of the solar nebula.
We  advocate averaging of multiple samples, rather than anchoring to individual meteorites, to improve precision.
}
\end{abstract}


\begin{highlights}
\item We re-evaluate and average data from the Al-Mg and Pb-Pb radiometric dating systems, for achondrites and chondrules, to attain better accuracy and precision in the time of formation of meteorites and meteoritic inclusions.
\item We find remarkable concordancy between the systems, provided the Pb-Pb age of the Solar System is $4568.4 \pm 0.2$ Myr, older than previous estimates; this suggests late resetting of the Pb-Pb system in Ca-rich, Al-rich inclusions (CAIs), and supports homogeneity of ${}^{26}{\rm Al}$ in the solar nebula.
\end{highlights}


\end{frontmatter}


%\begin{linenumbers}
% I. INTRODUCTION 
\section{Introduction} \label{sec:intro}

\subsection{The Al-Mg and Pb-Pb Chronometers}

To learn about the birth of planets and the story of our Solar System’s first few million years, we study meteorites that bear witness to this era. 
It is especially crucial to constrain the times at which their constituent components formed, the times at which their parent planetesimals accreted and melted, when collisions occurred, and to constrain the relative order of these events in the context of the solar nebula. 
The {\Steve need of astrophysical models is to know} the time $\Delta t$ after $t\!\!=\!\!0$ that an event occurred, where $t\!\!=\!\!0$ is a defined event or time in Solar System history. 

To obtain these times $\Delta t$, radiometric dating systems are employed. 
The Al-Mg system is most commonly used.
Calcium-rich, aluminum-rich inclusions (CAIs), incorporated live $\altwosix$, a short-lived radionuclide (SLR) that decays to ${}^{26}{\rm Mg}$ with a half-life of 0.717 Myr, or mean-life $\tau_{26} = 1.034$ Myr \citep{AuerEtal2009, Kondev2021}.
CAIs are among the first meteoritic components  to have formed in the Solar System, {\Steve apparently} during a short time interval that can be associated with $t\!\!=\!\!0$. 
Most CAIs incorporated Al with an isotopic ratio $\alratio \approx 
% (\alratio)_{\rm SS} = 
(5.23 \pm 0.13) \times 10^{-5}$ \citep{JacobsenEtal2008}.
{\Steve A useful working hypothesis is}
that this was the widespread and spatially homogeneous abundance of $\altwosix$ in the solar nebula when CAIs formed.  
We define $t\!\!=\!\!0$ to be the time in the solar nebula when $\alratio$ equalled $(\alratio)_{\rm SS} \equiv 5.23 \times 10^{-5}$ exactly.
Regardless of the ``true" initial value, most CAIs appear to have formed in a short time interval {\it around} this time \citep{LiuEtal2019}, so this is a convenient definition.
Assuming homogeneity of $\altwosix$, if a different object formed with a lower initial ratio $(\alratio)_0$, it must have formed later, at a time 
\begin{equation}
\Delta t_{26} = \tau_{26} \, \ln \left( \frac{ (\alratio)_{\rm SS} }{ (\alratio)_{0} } \right).
\end{equation}
This measures a time of formation relative to $t\!\!=\!\!0$.
{\Steve In practice, the Al-Mg chronometer often has uncertainty of less than $\pm 0.1$ Myr (\S 2).}

None of the original $\altwosix$ in CAIs exists today, and so its one-time existence must be inferred. 
Its initial abundance relative to stable Al is determined from a correlation between the isotopic ratio $({}^{26}{\rm Mg}/{}^{24}{\rm Mg})$ and the ratio $({}^{27}{\rm Al}/{}^{24}{\rm Mg})$, as measured in different minerals that {\Steve condensed or} crystallized from the same isotopic reservoir within the inclusion (or igneous achondrite). 
If the minerals have not been disturbed, e.g., not heated or altered in a way that would lead to {\Steve mobilization} of Mg isotopes, then the correlation must be linear and form an ``internal isochron" in a plot of $({}^{26}{\rm Mg}/{}^{24}{\rm Mg})$ versus $({}^{27}{\rm Al}/{}^{24}{\rm Mg})$ ratios of the inclusion’s minerals. 
The slope of the isochron is the inferred initial $(\alratio)_0$ ratio at the time the inclusion achieved isotopic closure.

Because $\altwosix$ must be inferred from an isochron, there are caveats associated with its determination.
First, determination of $(\alratio)_0$ is only possible if the data form a linear isochron.
This requires a statistical test: the mean squares weighted deviation (MSWD)] of the linear correlation must not exceed a maximum value $\approx 1 + 2 (2 / (N-2))^{1/2}$, where $N$ is the number of data points \citep{WendtCarl1991}.
A non-linear fit suggests disturbance or an additional process; without a model of this process, $(\alratio)_0$ cannot be inferred. 
Second, it must be remembered that the initial ratio $(\alratio)_0$ records not the formation of the CAI, but the time at which the Al and Mg isotopes no longer {\Steve moved} through the inclusion (i.e., when it reached isotopic closure). 
{\Steve On the parent body, aqueous alteration is a likely cause of disturbance. On the parent body or in the solar nebula, a} common cause of disturbance is heat-related diffusion of isotopes.
Often $(\alratio)_0$ records the time at which the CAI remained below a critical temperature, the ``closure temperature," such that diffusion of Mg isotopes is slower than continued cooling.
This is discussed further in (\S~\ref{sec:resetting}).

%Similar to the Al-Mg chronometer, the SLR $\mnfivethree$, which decays to ${}^{53}{\rm Cr}$ with a half-life of $3.7 \pm 0.4 (1\sigma)$ Myr \citep{HondaImamura1971}, can be used to radiometrically date times of formation of meteoritic components. 
%If the initial ratio $(\mnratio)_0$ is measured in a sample, and the solar nebula ratio $(\mnratio)_{\rm SS}$ at $t\!\!=\!\!0$ is known, then the sample formed at a time
%\begin{equation}
%\Delta t_{53} = \tau_{53} \, \ln \left( \frac{ ({}^{53}{\rm Mn}/{}^{55}{\rm Mn})_{\rm SS} }{ ({}^{53}{\rm Mn}/{}^{55}{\rm Mn})_{0} } \right)
%\end{equation}
%after $t\!\!=\!\!0$, where $\tau_{53}$ is the mean-life of $\mnfivethree$. 
%Other SLRs used for chronometry include ${}^{182}{\rm Hf}$ and ${}^{129}{\rm I}$.

%The Al-Mg and Mn-Cr isotopic systems are {\it relative} chronometers that do not give information on the absolute time of formation of an inclusion, except that it formed long enough ago that the SLR was abundant at the time but is now completely extinct.
%Instead, they provide information on the difference in time between two events in the solar nebula (e.g., formation of a CAI and a chondrule).
%This is the more important information for astrophysical models, allowing the order of events in the few Myr lifetime of the protoplanetary disk to be distinguished, to unrivaled precision. 
%The Al-Mg chronometer often has uncertainty of less than $\pm 0.1$ Myr.

{\Steve The Pb-Pb system provides a second way to measure times of formation.}
%The U-Pb system is thought of as an {\it absolute} chronometer, and indeed no other chronometer provides better absolute ages; but it is more useful as a relative chronometer.
% Ignoring the small fraction of atoms (e.g., ${}^{222}{\rm Rn}$) in the decay chain that may be lost by diffusion along the way,
In this system, ${}^{235}{\rm U}$ essentially decays to ${}^{207}{\rm Pb}$ with half-life $t_{1/2} = 703.81 \pm 0.96 (1\sigma)$ Myr, and ${}^{238}{\rm U}$ essentially decays to ${}^{206}{\rm Pb}$ with half-life $t_{1/2} = 4468.3 \pm 4.8 (1\sigma)$ Myr \citep{JaffeyEtal1971,VillaEtal2016}, but no natural isotopes decay to ${}^{204}{\rm Pb}$.
The ratio of radiogenic ${}^{207}{\rm Pb}$ to radiogenic ${}^{206}{\rm Pb}$ is therefore
\begin{equation}
\left( \frac{ {}^{207}{\rm Pb} }{ {}^{206}{\rm Pb} } \right)_{\rm r} = \left( \frac{ {}^{235}{\rm U} }{ {}^{238}{\rm U} } \right)_{\rm t} \, \frac{ \exp (+t/\tau_{235}) - 1 }{ \exp(+t/\tau_{238}) - 1 },
\label{eq:inverseisochron}
\end{equation}
where $({}^{235}{\rm U}/{}^{238}{\rm U})_{\rm t}$ is the isotope ratio measured in the sample today, and $t$ is the time that has elapsed since the inclusion formed.
The radiogenic Pb isotopic ratio is found by creating different washes and leachates and residues by  dissolution of a sample by different acids, and then measuring Pb isotopes in each.
A regression of $y={}^{207}{\rm Pb}/{}^{206}{\rm Pb}$ vs. $x={}^{204}{\rm Pb}/{}^{206}{\rm Pb}$ in each fraction yields a line (sometimes called the ``inverse isochron") with $y$-intercept (limit of zero non-radiogenic component) equal to the left-hand side of Equation~\ref{eq:inverseisochron}. 
Different washes or leachates may incorporate variable amounts of Pb from terrestrial contamination or initial non-radiogenic Pb, either of which may cause a particular data point to fall off an otherwise reasonable regression.
As with Al-Mg isochrons, adherence to a line can be tested using the MSWD of the linear regression.
Before about 2010, it was assumed that all samples were characterized by $\uratio = 137.88$; but it has become recognized that CAIs, achondrites, etc., vary significantly in this ratio.
{\Steve This is attributable to variable amounts of radiogenic ${}^{235}{\rm U}$ from decay of ${}^{247}{\rm Cm}$ \citep{BrenneckaEtal2010,TissotEtal2016}, but also evaporation and other processes \citep{TissotDauphas2015,TissotEtal2017,TissotEtal2019}.}
A fractional change of $10^{-3}$ in the U isotope ratio corresponds to a shift in the age of 1.45 Myr, {\Steve and typical variations in CAIs can lead to inaccuracies in their ages $> 1$ Myr.}
Therefore it is now standard to report ``U-corrected" Pb-Pb ages, in which ${}^{238}{\rm U}/{}^{235}{\rm U}$ is measured in the sample.

%{\Steve
%For samples for which the intercept $r = ({}^{207}{\rm Pb}/{}^{206}{\rm Pb})_{\rm r}$ is measured from an isochron in the sample, and for which $u$ is measured {\bf in the same sample}, two useful relations are 
%\begin{equation}
%t_{\rm Pb} \approx \left[ 3120.243 + 16.791 \, u \, r \right] \, {\rm Myr}
%t_{\rm Pb} \approx \left[ 3119.914 + 16.7856 \, u \, r \right] \, {\rm Myr}
%\label{eq:PbPbage}
%\end{equation}
%and
%\begin{equation}
%\delta t_{\rm Pb} \approx 1450 \, \left[ \left( %\frac{\delta u}{u} \right)^2 + \left( %\frac{\delta r}{r} \right)^{2} \right]^{1/2} \, %{\rm Myr},
%\label{eq:PbPberror}
%\end{equation}
%where it is assumed that the samples have ages %$\approx 4568$ Myr, and the standard half-lives %of ${}^{235}{\rm U}$ and ${}^{238}{\rm U}$ %(above) are assumed.
%Representative values might be $u \approx 137.8$ and $\delta u \approx 0.01$ and $r \approx 0.625$ and $\delta r \approx 0.0001$, yielding $t_{\rm Pb} \approx 4568.4$ Myr and $\delta t_{\rm Pb} \approx 0.25$ Myr.
%This reflects the measurement uncertainty in the U isotope ratios and the uncertainty in the intercept (given a set of data that are regressed), ultimately also due to measurement errors.
%It does not reflect other sources of systematic errors.
%}

{\EDIT An additional issue is that not all uranium isotopic ratios are reported using the same assumed values for standards. 
This may lead to variations in ${}^{238}{\rm U}/{}^{235}{\rm U}$ on the order of $0.05\permil$, leading to variations in inferred Pb-Pb ages of 0.07 Myr (F. Tissot, personal communication). 
Future renormalization of the data appears warranted.}

The uncertainty in Pb-Pb absolute ages is often misunderstood to be $< 1$ Myr.
{\Steve (i.e., $\pm y$ in the notation of \citet{TissotEtal2017}), but ignores other sources of uncertainty, the largest of which is the half-lives of U. 
Although both are known at the 0.1\% level,}
the uncertainty in absolute Pb-Pb ages is $\pm 9$ Myr, due to uncertainties in the ${}^{238}{\rm U}$ and especially ${}^{235}{\rm U}$ half-lives \citep{Ludwig2000,TissotEtal2017}.
However, when two Pb-Pb ages are computed using the same half-lives, the systematic uncertainties cancel, and the ($2\sigma$) precision can approach $0.3 - 0.5$ Myr \citep{TissotEtal2017}. 
{\Steve It is essential that the half-lives be identical.}
Other half-lives are sometimes used---e.g., the ${}^{235}{\rm U}$ half-life of 703.20 Myr \citep{Mattinson2010} used by \citet{PalkEtal2018} in their Pb-Pb dating of enstatite chondrites---but fortunately almost all absolute ages in the literature are computed using the half-lives quoted above from \citet{JaffeyEtal1971} and \citet{VillaEtal2016}.
{\Steve
The Pb-Pb system is the most precise and versatile absolute chronometer, but its greater utility and precision comes from using it as a {\it relative} chronometer to determine the sequence of events in the solar nebula. 
}

To be effective as a relative chronometer, absolute ages measured by the Pb-Pb system, $t_{\rm Pb}$, must be converted into times after $t\!\!=\!\!0$, $\Delta t_{\rm Pb}$, by subtracting them from the absolute age of $t\!\!=\!\!0$, which we denote $t_{\rm SS}$:
\begin{equation}
\Delta t_{\rm Pb} = t_{\rm SS} - t_{\rm Pb}.
\label{eq:tCAI}
\end{equation}
Here $t_{\rm SS}$ should be interpreted as the Pb-Pb age that would be obtained by measuring the age of a sample that achieved isotopic closure at $t\!\!=\!\!0$, using the same uranium half-lives as assumed for other samples.
Determination of $t_{\rm SS}$, the ``age of the Solar System," is {\Steve one focus of this paper}.

{\Steve
\subsection{Absolute Age of CAIs and the Solar System}
}

To date, most determinations of $t_{\rm SS}$ and the time elapsed since $t\!\!=\!\!0$ have been made by radiometrically dating CAIs. 
As described above, an inverse isochron is made by  measuring Pb isotopes in different washes, leachates and residues (``fractions") obtained by reacting a sample with various acids, and for each fraction obtaining isotopic fractionation-corrected $y={}^{207}{\rm Pb}/{}^{206}{\rm Pb}$ and $x={}^{204}{\rm Pb}/{}^{206}{\rm Pb}$ ratios.
Assuming a linear correlation, the $y$-intercept of the line is the purely radiogenic end-member $({}^{207}{\rm Pb}/{}^{206}{\rm Pb})$, and this plus a measurement of ${}^{235}{\rm U}/{}^{238}{\rm U}$ allows one to solve for $t$ in Equation~\ref{eq:inverseisochron}.
The uncertainty in the age comes from adding in quadrature the uncertainties in the ${}^{235}{\rm U}/{}^{238}{\rm U}$ ratio and the uncertainties in the $y$-intercept from the linear regression.
The latter derives from the measurement uncertainties in the $x$ and $y$ data points.

{\Steve Perhaps because}
of the difficulty of the measurement (wet chemistry of small samples while avoiding terrestrial contamination) and the newness of the uranium correction, there are only four CAIs with peer-reviewed, U-corrected Pb-Pb ages.
\citet{AmelinEtal2010} dated Allende CAI \textit{SJ101} at $4567.18 \pm 0.50$ Myr.
%\citet{AmelinEtal2010} found a Pb-Pb age for Allende CAI \textit{SJ101} of $4567.18 \pm 0.50$ Myr from an isochron with MSWD $= 1.07$, using a subset of the fractions and a weighted average of others. 
{\Steve
\citet{MacPhersonEtal2017} reported an internal isochron with $(\alratio)_0 = (5.20 \pm 0.53) \times 10^{-5}$ for this CAI.}
%implying $\Delta t_{26} = 0.006_{-0.100}^{+0.111}$ Myr. 
%If both the Pb-Pb and Al-Mg systems achieved closure at the same time, this would imply $\tCAI = 4567.19 \pm 0.51$ Myr. 
\citet{ConnellyEtal2012} found ages for Efremovka CAIs \textit{22E}, \textit{31E} and \textit{32E} of $4567.35 \pm 0.28$ Myr, $4567.23 \pm 0.29$ Myr, and $4567.38 \pm 0.21$ Myr. 
{\Steve \citet{LarsenEtal2011} found excesses of ${}^{26}{\rm Mg}$ in \textit{22E} and \textit{31E}
that are consistent with formation with $(\alratio)_0 \approx 5.23 \times 10^{-5}$.}
%\citet{ConnellyEtal2012} found an age of Efremovka CAI \textit{22E} of $4567.35 \pm 0.28$ Myr from an isochron with MSWD $= 0.91$, using 11 of 20 fractions in the regression; an age for Efremovka CAI \textit{31E} of $4567.23 \pm 0.29$ Myr from an isochron with MSWD $=1.19$, using 5 of 8 fractions, plus a weighted average of 9 other fractions (effectively 6 data points out of a possible 17) in the regression; and an age for Efremovka CAI \textit{32E} of $4567.38 \pm 0.31$ Myr from an isochron with MSWD $=0.49$, using 6 of 11 fractions in the regression. 
%The weighted mean of these values is $4567.32 \pm 0.17$ Myr.
{\Steve {\it Assuming} that all of these CAIs experienced isotopic closure in Pb-Pb at {\EDIT identical times}, that time is the weighted average of the Pb-Pb ages, given by \citet{ConnellyEtal2012} as $4567.30 \pm 0.16$ Myr. 
{\it Further assuming} that they all closed in Al-Mg at this
 same time,}
%Because \citet{MacPhersonEtal2017} and \citet{LarsenEtal2011} found that CAIs \textit{SJ101}, \textit{22E}, and \textit{31E} appear to have canonical $(\alratio)_0$, 
this time is taken to be $t\!\!=\!\!0$.
%They then averaged this with the value for \textit{SJ101} to recommend a single value $4567.30 \pm 0.16$ Myr, which implicitly assumes that all 4 CAIs achieved isotopic closure simultaneously.
%\citet{LarsenEtal2011} reported 
%$(\alratio)_0 = (5.25 \pm 0.02) \times 10^{-5}$ for CAI \textit{22E} (based on a bulk CAI value, not an internal isochron), implying $\Delta t_{26} = -0.004$ up to $0.000$ Myr.
%If applicable to the other three CAIs, this also would imply $\tCAI = 4567.30 \pm 0.16$ Myr.
These four CAIs are the only ones with peer-reviewed, U-corrected Pb-Pb ages, and all appear consistent with this average age, so this value has been widely adopted as the time elapsed since $t\!\!=\!\!0$. 

There are strong hints of older ages of CAIs, however.
\citet{BouvierEtal2011a} reported a U-corrected Pb-Pb age for NWA 6991 CAI {\it B4} of $4567.94 \pm 0.31$ Myr.
%on an isochron with MSWD $= 1.8$, using a regression with 5 fractions and not rejecting any.
% [The maximum acceptable MSWD for $N=5$ is 2.6.]
%As discussed by \citet{SanbornEtal2019}, the true uncertainty on this age is probably $\pm 0.43$ Myr after propagating the uncertainties in the U isotopic ratio. 
This CAI also has canonical $(\alratio)_0$ \citep{WadhwaEtal2014}.
%\citet{WadhwaEtal2014} reported $(\alratio)_0 = (4.90 \pm 0.05) \times 10^{-5}$ for this CAI, implying $\Delta t_{26} = 0.067 \pm 0.010$ Myr and therefore $\tCAI = 4568.01 \pm 0.43$ Myr.
In addition, \citet{BouvierWadhwa2010} inferred an age $4568.22 \pm 0.18$ Myr in NWA 2364 CAI {\it B1}.
These ages have not been widely accepted, perhaps because the former has not been published in the refereed literature, and the latter was not U-corrected using a direct measurement of U isotopes, but rather a correlation between Th/U and $\uratio$ found by \citet{BrenneckaEtal2010};
{\Steve this relation is not universal \citep{TissotEtal2016}, and so might} not yield the correct age adjustment for this CAI. 
The two latter ages do not agree with $4567.30 \pm 0.16$ Myr to within the uncertainties. 
If accepted or confirmed, CAIs \textit{B1} and \textit{SJ101} would be seen to have formed at times that differ {\Steve from $4567.3$ Myr by over} 1.0 Myr (at the $4\sigma$ level).

More Pb-Pb dates of CAIs would be enormously helpful, to determine whether CAIs reached isotopic closure in the Pb-Pb system at the same time or across a range of times. 
Because the four CAIs averaged by \citet{ConnellyEtal2012} seem to have formed at a single time, that time is taken to be $t\!\!=\!\!0$. 
However, as is well-known and as is discussed below, if CAIs actually formed with canonical $(\alratio)_0 \approx 5.23 \times 10^{-5}$ at 4567.30 Myr ago, then almost all other chronometers would be left discordant. 
This  has led {\Steve to the proposal} that $\altwosix$ must have been spatially or temporally heterogeneous in the solar nebula, with the CAI-forming region holding nearly four times as much $\altwosix$ as the rest of the solar nebula \citep{LarsenEtal2011,BollardEtal2019}.

{\Steve In this paper},
%In \S 4.1, 
we explore other reasons why Al-Mg and Pb-Pb ages might not match.
One {\Steve important factor we identify} is that the practice of building isochrons from a subset of fractions means that {\EDIT some} Pb-Pb ages are inaccurate and their uncertainties have been considerably underestimated. 
Another {\Steve pertinent fact} is that CAIs were {\Steve likely} transiently heated at random times {\EDIT up to 3} Myr after $t\!\!=\!\!0$; {\Steve we demonstrate} that the Pb-Pb system {\Steve could have been} isotopically reset, without disturbing the Al-Mg system. 
{\Steve As these are plausible scenarios, it is a distinct possibility that the reported Pb-Pb ages of CAIs might not actually record the time of $t\!\!=\!\!0$.}

{\Steve 
An alternative to using radiometric dating of CAIs to determine $t_{\rm SS}$ is to mathematically find values that reduce the discrepancies between the Pb-Pb and other chronometers to levels consistent with measurement errors. 
Many other groups have sought to correlate Al-Mg and other chronometers. 
For example,
\citet{LugmairShukolyukov1998} correlated Mn-Cr ages against Pb-Pb ages in many achondrites, finding a range of ages 4568 to 4571 Myr would minimize the discrepancies between the chronometers. 
These ages overlapped with the then-accepted age of CAIs, $4566 \pm 2$ Myr \citep{GopelEtal1991}, so it was possible to conclude the chronometers were concordant.
Of course, none of these ages was U-corrected.
\citet{NyquistEtal2009} correlated Al-Mg and Mn-Cr ages to show that their ages were linearly correlated and thereby constrain the half-life of $\mnfivethree$ and the Solar System initial value $(\mnratio)_{\rm SS}$.
They then used the initial $(\mnratio)_0$ inferred for the achondrite LEW 86010 to determine the time of its formation after $t\!\!=\!\!0$. 
They added this to the Pb-Pb age of LEW 86010 to derive $t_{\rm SS} = 4568.2$ Myr.
This Pb-Pb age was not U-corrected, either.
%That and other factors probably precluded widespread adoption of this value as the age of the Solar System. 
\citet{SanbornEtal2019} directly correlated U-corrected Pb-Pb ages against Al-Mg formation times for several achondrites and CAIs. 
Although not explicitly stated, their Figure 6 makes clear that they would infer a value of $t_{\rm SS} \approx 4567.8$ Myr, with an uncertainty that we estimate at $\sim 0.5$ Myr.
Recently, \citet{PirallaEtal2023} also directly correlated U-corrected Pb-Pb ages against Al-Mg formation times of several achondrites and concluded $t_{\rm SS} = 4568.7$ Myr.
We further discuss these treatments below.
One key aspect missing from these treatments is that they did not check whether the values of $t_{\rm SS}$ they derived led to a statistically valid fit between the Al-Mg and other chronometers, i.e., whether the discrepancies between their inferred $\Delta t_{26}$ and other formation times are attributable solely to measurement uncertainties.
Without this assessment of concordancy, it is impossible to test the underlying assumption of $\altwosix$ homogeneity.
}
\vspace{0.1in}

%These analyses that show such a large spread in CAI Pb-Pb ages raises the question of whether Al-Mg and U-Pb systems closed simultaneously in them and whether they are suitable for such analyses. 
%As well, since these anlayses were done, new data from chondrules have been produced that could be included \citep{BollardEtal2019}, as well as multiple datasets for achondrites that could be vetted and averaged together.

\subsection{Outline}

{\Steve
In this work we test the hypothesis that $\altwosix$ was distributed homogeneously in the solar nebula.
We do this by comparing Al-Mg and Pb-Pb ages in the samples above, as in the work by \citet{SanbornEtal2019}.
A prediction of the hypothesis is that in samples in which the systems must have closed simultaneously, a single value $t_{\rm SS}$ must reconcile the chronometers, in a statistical sense. 
If a single value of $t_{\rm SS}$ does not reconcile the Al-Mg and Pb-Pb ages, then the hypothesis of $\altwosix$ homogeneity is falsified.
If the chronometers {\it are} reconciled, then that supports the homogeneous hypothesis, and futher suggests that the range of values of $t_{\rm SS}$ that reconciles them is the Pb-Pb age of the Solar System at $t\!\!=\!\!0$.}

{\Steve 
In \S 2 we compile and discuss the meteoritic data, which comprises five achondrites from the non-carbonaceous chondrite (NC) isotopic reservoir, two achondrites from the carbonaceous chondrite (CC) isotopic reservoir, four chondrules, and one CAI. 
We review closure temperatures and why there can be no discrepancies between Al-Mg and Pb-Pb ages for the rapidly cooled achondrites, and therefore why these can falsify the homogeneous hypothesis. 
We discuss why the other samples, especially the CAI, do not necessarily test the hypothesis.
We re-analyze the Pb-Pb isochrons from all samples, to determine the Pb-Pb age and the uncertainty for each. 
}

{\Steve
In \S 3 we take weighted means of the data to derive the value of $t_{\rm SS}$ that minimizes the discrepancies between the Al-Mg and Pb-Pb chronometers in the appropriate samples.
We calculate the goodness-of-fit metric $\chi_{\nu}^{2}$ (also known as MSWD) to test whether the samples are adequately fit by a single value of $t_{\rm SS}$.
We find the the Pb-Pb and Al-Mg ages are indeed reconciled in a statistical sense for the seven achondrites, and even the 4 chondrules as well, 
%(quenched angrites D'Orbigny, SAH 99555, NWA 1670; pseudo-eucrite Asuka 881394; basalatic achondrite NWA 7325), 
if $t_{\rm SS} = 4568.42 \pm 0.24$ Myr.
Had they not been, this would have falsified the hypothesis of homogeneity, but instead this finding supports it.
We find that the same $t_{\rm SS}$ that minimizes the discrepancies between the Al-Mg and Pb-Pb systems among the NC achondrites also makes the chondrules concordant, but cannot reconcile the {\Steve CAI {\it SJ101}}, which appears reset.
}

% We also consider the fit of four NWA 5697 chondrules and the two carbonaceous achondrites NWA 2976 and NWA 6704.
% Adding the four chondrules that have been U-corrected using a measured ${}^{238}{\rm U}/{}^{235}{\rm U}$, we find an essentially identical $t_{\rm SS} = 4568.68 \pm 0.15$, and all nine samples are consistent with that value at the $< 1.7\sigma$ level. 
% We find that the `carbonaceous achondrites' NWA 6704 and NWA 2796 are not consistent with this value, strongly suggesting that the Al-Mg and Pb-Pb systems did not close simultaneously in them.

{\Steve 
In \S 4 we compare our approach to previous similar approaches.
We discuss how the data support homogeneity of $\altwosix$ and discuss why astrophysical models strongly support such an interpretation.
One of our primary conclusions is that the Pb-Pb age of CAI \textit{SJ101} and other CAIs may have been reset 1.2 Myr after it formed and recorded an Al-Mg age, which was not reset.
We strongly support reporting times of formation after $t\!\!=\!\!0$ rather than absolute ages. 
Astrophysical models are unable to use {\Steve absolute ages}, only relative formation times, and relative times of formation are far more precise anyway.
}

%In \S 4 we consider why Al-Mg and Pb-Pb might not have closed simultaneously in the carbonaceous achondrites.
%We consider it likely that they probably achieved isotopic closure in the Pb-Pb system at lower temperatures, characterized by slower cooling, compared to non-carbonaceous achondrites. 
%Likewise, we consider why our predicted $t_{\rm SS}$ is 1.4 Myr older than the Pb-Pb age inferred from measurements of CAIs that formed with canonical $(\alratio)_0$ at $t\!\!=\!\!0$.
%We consider it likely that CAIs were transiently heated, consistent with the peak temperatures and cooling rates experienced by type B CAIs, at late times in the protoplanetary disk, and that we show that this could have reset the Pb-Pb system without disturbing the Al-Mg system. 

{\Steve
In \S 5 we draw conclusions.
It does not appear that  $\altwosix$ was heterogeneous in the solar nebula.
}
We argue that statistical approaches, like those of \citet{NyquistEtal2009}, \citet{SanbornEtal2019}, \citet{PirallaEtal2023}, and presented here, are more reliable determinants of $t_{\rm SS}$ than direct measurements of CAIs.
We advocate for reporting dates as times after $t\!\!=\!\!0$, not absolute ages, and we advocate against use of individual anchors.

\section{Meteoritic Data}

{\Steve

\subsection{Which Samples test Homogeneity?}

The hypothesis of homogeneous $\altwosix$ makes the testable prediction that times of formation of appropriate samples found using the Al-Mg chronometer should match times of formation found using the Pb-Pb chronometer.
Several conditions must be met for a sample to be an appropriate test.
The most important is that the $(\alratio)_0$ and Pb-Pb age actually can be found in an object; the former requires a valid Al-Mg isochron, and the latter a valid Pb-Pb isochron and measurement of the ${}^{238}{\rm U}/{}^{235}{\rm U}$ ratio in the sample. 
It also must be {\bf certain} that both the Al-Mg and Pb-Pb systems date the {\bf same} event.
The Al-Mg ages can be defined as a time of formation after $t\!\!=\!\!0$, $\Delta t_{26}$, using Equation 1.
The Pb-Pb absolute ages must be converted to a time of formation after $t\!\!=\!\!0$, $\Delta t_{\rm Pb}$, using Equation 3. 

These requirements drive our choice of samples.
To our knowledge, the objects in which both Al-Mg and U-corrected Pb-Pb ages have been determined includes seven bulk achondrites, four chondrules, and one CAI.
The achondrites include five with the signatures of having formed in the NC isotopic reservoir:
the quenched (volcanic) angrites D'Orbigny, SAH 99555, and NWA 1670; the eucrite-like achondrite Asuka 881394; and the ungrouped achondrite NWA 7325 (paired with NWA 8486).
Two of the achondrites have signatures of having formed in the CC isotopic reservoir: the ungrouped achondrites NWA 2976 (paired with NWA 011) and NWA 6704 (paired with NWA 6693).
In some samples, we are confident that the Al-Mg and Pb-Pb systems probably date the same event; in others we have no evidence to support that.

There are many reasons why the two isotopic systems might not date the same event. 
As discussed below, bulk achondrites generally record the time at which the temperature dropped below a ``closure temperature".
This can be hundreds of K different for different minerals and isotopic systems, so Al-Mg and Pb-Pb {\Steve can} record different times. 
If the cooling rate is slow, only $\sim 10^2 \, {\rm K} \, {\rm Myr}^{-1}$, then those times can be $\sim 10^6$ years apart.
Alternatively, chondrules notoriously were exposed to transient heating events with peak temperatures $> 1800$ K, followed by cooling rates $\sim 10 - 10^3 \, {\rm K} \, {\rm hr}^{-1}$ \citep{DeschEtal2012}.
Being located in the nebula at the same time as chondrules before parent-body accretion, CAIs also could have been exposed to these same heating events.
Below we explore how such transient heating events could have reset the Pb-Pb isochron at a late time, {\Steve and even do so} without resetting the Al-Mg chronometer, which would record a much earlier event.
}

{\Steve
\subsection{Closure Temperatues of the Al-Mg and Pb-Pb Systems}
}
\label{sec:resetting}

{\Steve
To quantify the largest possible difference between the times recorded by the Al-Mg and Pb-Pb systems, we calculate their closure temperatures,
the temperatures below which isotopic redistribution within the sample (here, by thermal diffusion) is slower than further cooling.}
Closure temperatures are found by comparing the cooling rate of the sample against the diffusion rate of key elements in specific minerals, and depend on the grain size of the minerals
 {\EDIT (more precisely, the lengthscales over which isotopic gradients must be preserved).}
It is assumed the diffusion coefficient of the element obeys an Arrhenius relationship:
\begin{equation}
D(T) = D_0 \, \exp \left( -E / R T \right),
\end{equation}
where $D_0$ is the pre-factor and $E$ the activation energy, both specific to the element and mineral, and $R$ the gas constant. 
In that case the closure temperature is given by
\begin{equation}
T_{\rm c} = \frac{E}{R} \, \left[ 
 \ln \left( \frac{ A \, T_{\rm c}^2 \, D_0 }{ (E / R) \, (dT/dt) \, a^2 } \right) \right]^{-1} 
\end{equation}
\citep{Dodson1973},
where $A=55$ 
{\Steve (for a sphere; $A=27$ for a cylinder, $A=8.7$ for a plane)},
$a$ is the grain radius, $dT/dt$ is the cooling rate, and it is assumed that the cooling is from a peak temperature above $T_{\rm c}$.
The closure temperature must be estimated, substituted into the right-hand side, and the solution iterated to convergence.
The majority of Pb-Pb dates are based on data from pyroxene grains.
We have calculated the closure temperature for Pb in clinopyroxene using 
$D_0 = 44 \, {\rm m}^{2} \, {\rm s}^{-1}$, and $E / R = 62,400$ K \citep{Cherniak1998}.
Al-Mg measurements have been taken in a variety of minerals, including anorthite, melilite, pyroxene and spinel.
For Mg diffusion in anorthite, we assume $D_0 = 1.2 \times 10^{-6} \, {\rm m}^{2} \, {\rm s}^{-1}$ and 
$E / R = 33440$ K \citep{LaTouretteWasserburg1998}.
For Mg diffusion in melilite, we assume $D_0 = 3.02 \times 10^{-9} \, {\rm m}^{2} \, {\rm s}^{-1}$ and $E / R = 28990$ K \citep{ItoGanguly2009}. 
For Mg diffusion in pyroxene, we assume $D_0 = 1.9 \times 10^{-9} \, {\rm m}^{2} \, {\rm s}^{-1}$ and $E / R = 24300$ K \citep{MullerEtal2013}.
For Mg diffusion in spinel, we assume $D_0 = 2.77 \times 10^{-7} \, {\rm m}^{2} \, {\rm s}^{-1}$ and $E / R = 38560$ K \citep{LiermannGanguly2002}.

\begin{center}
\begin{figure}[ht!]
\includegraphics[width=0.99\textwidth,angle=0]{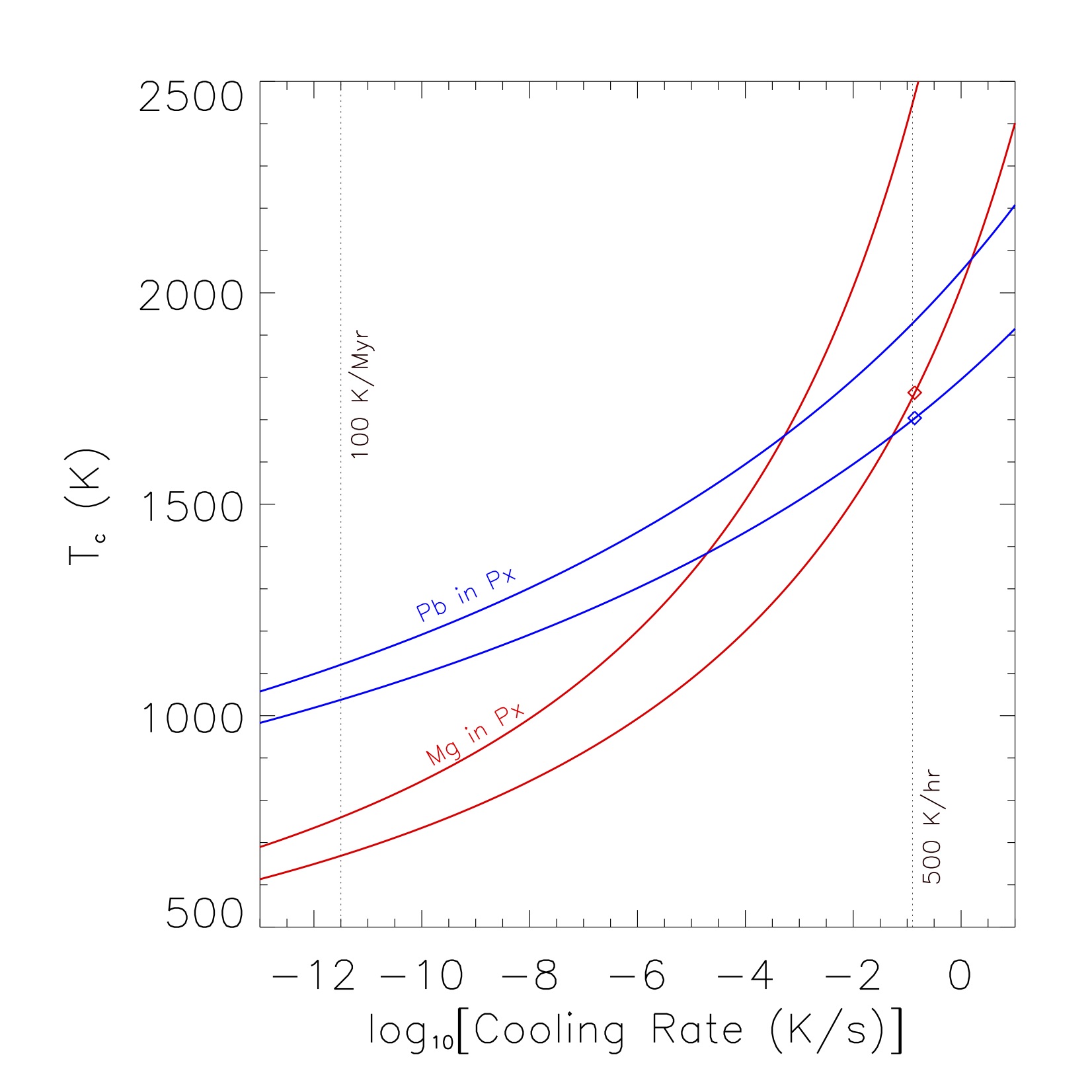}
\caption{
Closure temperatures of Pb in clinopyroxene (blue curves) and Mg in clinopyroxene (red curves), as a function of cooling rate, assuming grain sizes (diameters) of $200 \, \mu{\rm m}$ (top), and $20 \, \mu{\rm m}$ (bottom). For slow cooling rates typical of secular cooling on asteroids (e.g., 100 K/Myr), Al-Mg has a lower closure temperature than the Pb-Pb system and is more easily reset. For a cooling rate 500 K/hr, associated with transient heating events experienced by CAIs or chondrules, the closure temperatures are reversed. At a hypothetical peak temperature of $\approx 1730$ K and cooling rate 500 K/hr, the Al-Mg system would remain closed in clinopyroxene grains $100 \, \mu{\rm m}$ in size (closure temperature 1764 K, red diamond), but the Pb-Pb system in pyroxene grains $100 \, \mu{\rm m}$ in size would be disturbed (closure temperature 1704 K, blue diamond). 
\label{fig:diffusion}}
\end{figure}
\end{center}

In general, during transient heating events, Pb-Pb is easier to reset than Al-Mg.
In {\bf Figure~\ref{fig:diffusion}} we plot the calculated closure temperatures $T_{\rm c}$ of Pb and Mg in clinopyroxene, as functions of the cooling rate for grain sizes of 20 and $200 \, \mu{\rm m}$. 
At slow cooling rates characteristic of secular cooling of asteroids, $\sim 100 \, {\rm K} \, {\rm Myr}^{-1}$ ($3 \times 10^{-12} \, {\rm K} \, {\rm s}^{-1}$), $T_{\rm c} \approx 1040$ K for Pb, but lower ($\approx 670$ K) for Mg.
The order of isotopic closure tends to reverse for small inclusions in transient heating events with cooling rates $\sim 1 - 10^3 \, {\rm K} \, {\rm hr}^{-1}$ ($3 \times 10^{-4} - 0.3\, {\rm K} \, {\rm s}^{-1}$). 

%---------------------------------
% Figure 1
%---------------------------------

{\Steve
The quenched or ``volcanic" angrites (D'Orbigny, SAH 99555, NWA 1670) are widely accepted, based on petrologic features suggesting a lava erupted onto an asteroid surface, to have cooled very rapidly, at rates $\approx 300 \, {\rm K} \, {\rm hr}^{-1}$ in the 1273 - 1573 K range \citep{Keil2012}.
Likewise, NWA 7325 shares these petrologic features, and is inferred from Mg diffusion in plagioclase to have cooled at rates $\sim 500-650 \, {\rm K} \, {\rm hr}^{-1}$, at high temperatures \citep{YangEtal2019}.
%It may have been remelted \citep{WeberEtal2016}, but we assume the remelting reset all the chronometers. 
The achondrite NWA 6704 was inferred by \citet{HibiyaEtal2019} from textures to have cooled at rates $1 - 10^2 \, {\rm K} \, {\rm hr}^{-1}$ at high temperatures, and from olivine-spinel geospeedometry to have cooled at rates $< 10^{-2} \, {\rm K} \, {\rm hr}^{-1}$ at 1123 - 1243 K.
The cooling rate of Asuka 881394 has not been constrained petrologically; but given its early time of formation in common with the achondrites above, we assume it also cooled rapidly. 
It is very likely that these achondrites formed from complete melts at high temperatures and continued to cool below the closure temperatures of Al-Mg and Pb-Pb in a matter of years, so that these two systems closed effectively simultaneously. 
{\bf Al-Mg and (properly determined) Pb-Pb ages must be concordant in these achondrites.}
}

{\Steve
In contrast to the achondrites, the Al-Mg and Pb-Pb systems will not {\bf necessarily} have closed at the same time in chondrules and CAIs.
This is not because they were not rapidly cooled, but because their last heating event did not reset both chronometers.
Chondrules are silicate melts that crystallized following a transient heating event in the solar nebula, typified by peak temperatures anywhere from subsolidus or barely above the solidus \citep{WassonRubin2003,RuzickaEtal2008}, to $> 1800$ K or more, and cooling rates $\sim 10 - 10^3 \, {\rm K} \, {\rm hr}^{-1}$ \citep{DeschEtal2012}.
Many chondrules appear to have experienced multiple such events.
}
A representative thermal history might be heating to a peak temperature $\approx 1700-1800$ K, followed by cooling at $500 \, {\rm K} \, {\rm hr}^{-1}$ (a cooling rate consistent with all chondrule textures).

A majority of CAIs also were transiently heated in the solar nebula.
Compact type A CAIs were at least partially melted or annealed, and type B and C CAIs have igneous textures showing they were partially or completely melted at least once.
During their last heating event, type B CAIs appear from zoning in melilite to have cooled at rates of $0.5 - 50 \, {\rm K} \, {\rm hr}^{-1}$ \citep{StolperPaque1986,JonesEtal2000}, or (from the need to limit diffusion of oxygen isotopes) $> 6 \times 10^4 \, {\rm K} \, {\rm hr}^{-1}$ \citep{KawasakiEtal2021}.
These hours-long cooling times are of the same order as the cooling times experienced by chondrules.

Both chondrules and CAIs resided in the same regions of the solar nebula, for the same $\sim 2-4$ Myr it took before chondrites formed \citep{DeschEtal2018}. 
Given that a large fraction of the material in chondrites is chondrules, all of which were transiently heated by some event, it would be remarkable if CAIs avoided being transiently heated as well.
It seems very likely that CAIs must have been transiently heated, some to the melting point and some to only subsolidus temperatures, over the same period of time as chondrules (1-3 Myr after $t\!\!=\!\!0$; \citealt{VilleneuveEtal2009}).
{\Steve In fact, evidence exists of this late heating of CAIs \citep{ManeEtal2022}.}

{\Steve 
These transient heating events could have reset isotopic systems in chondrules and CAIs, depending on the peak temperatures and the cooling rates.}
If a CAI were to be heated to peak temperatures {\Steve of about 1750 K}, and cool at $500 \, {\rm K} \, {\rm hr}^{-1}$, pre-existing minerals would survive. 
Pyroxene (enstatite) grains survive to temperatures of 1830 K \citep{GreenwoodHess1996},
{\Steve
although the more Ti-rich pyroxene more typically found in CAIs may partially melt at lower temperatures $\approx 1500$ K (A. Davis, personal communication).
}
Spinel grains survive to similar temperatures $> 1800$ K \citep{WhattamEtal2022}.
{\Steve
For $20 \, \mu{\rm m}$ grains ($a = 10 \mu{\rm m}$) and cooling rates of 500 K/hr, we calculate closure temperatures of 2071 K in spinel, 2022 K in melilite, 1764 K in pyroxene, and 1687 K in anorthite.
With the possible exception of anorthite, which would be slightly disturbed, a heating event with peak temperatures just above 1700 K would not reset the Al-Mg chronometer.}
%The Al-Mg system has a closure temperature of 1764 K in clinopyroxene, and 2022 K in spinel, and would {\bf not} be reset, 
{\Steve
In contrast, the closure temperature of the Pb-Pb system in $20 \, \mu{\rm m}$ pyroxene grains would be no higher than 1703 K, so Pb-Pb ages would be reset, especially if the more Ti-rich pyroxene grains were to partially melt.
}
{ The types of transient heating events experienced by chondrules and assuredly also CAIs are capable of resetting the Pb-Pb chronometer without necessarily resetting the Al-Mg chronometer, at least if it Al-Mg ages are determined using isochrons based on pyroxene and spinel.
Chondrules and CAIs may have formed at $t\!\!=\!\!0$, or any time over the next 2-3 Myr, before a transient heating event reset their Pb-Pb ages without necessarily resetting their Al-Mg ages. 
Therefore chondrules and CAIs are not samples that reliably or necessarily test the homogeneity hypothesis.}

It has been noted before that the discrepancies between the Al-Mg and Pb-Pb systems imply a late-stage resetting of the Pb-Pb system.
\citet{BouvierWadhwa2010} suggested that diffusion of Pb in pyroxene might be faster than diffusion of Mg in melilite or anorthite, although they did not outline a scenario in which this would occur, identify a nebular or parent-body setting, or quantify the thermal histories that would be required. 

{\Steve
With these caveats, we examine the ages recorded by each sample.}

\subsection{Achondrites}

\subsubsection{D'Orbigny (quenched angrite)}

D'Orbigny is a quenched angrite with an unshocked and unbrecciated texture consisting of large laths of anorthite intergrown with Al-Ti-bearing diopside-hedenbergite, Ca-rich olivine and kirschsteinite as well as abundant glasses and round vugs \citep{Keil2012}. 
{\Steve 
The glass is not of shock origin or externally introduced at  a later time \citep{GlavinEtal2004}.
The glass and vugs speak to a rapid cooling rate $\sim 300 \, {\rm K} \, {\rm hr}^{-1}$ \citep{Keil2012}.
D'Orbigny has long been considered an anchor in which the different isotopic systems likely closed simultaneously.}

The initial $(\alratio)_0$ ratio in D'Orbigny has been determined by multiple groups:
\citet{SpivakBirndorfEtal2009} found $(5.06 \pm 0.92) \times 10^{-7}$, MSWD $=2.5$;
\citet{SchillerEtal2010} found $(3.88 \pm 0.27) \times 10^{-7}$, MSWD $=1.9$; and 
\citet{SchillerEtal2015} found $(3.98 \pm 0.15) \times 10^{-7}$, MSWD $=1.9$.
Combining these data and rejecting one outlier, \citet{SanbornEtal2019} found $(\alratio)_0 = (3.93 \pm 0.39) \times 10^{-7}$, the value we adopt. 
It has also been reported to have been $(3.97 \pm 0.21) \times 10^{-7}$, but from an isochron with MSWD above the threshold for acceptance \citep{KleineWadhwa2017}. 
For our adopted initial ratio
of $5.23 \times 10^{-5}$, we find a time of formation \mbox{\boldmath$\Delta t_{26} = 5.06 \pm 0.10 \, {\rm Myr}$}
(Equation 1).

\citet{Amelin2008a} measured Pb isotopic ratios using whole-rock and pyroxene fractions in D'Orbigny.  
Excluding two outliers with excess initial Pb, and one with low Pb concentration, and using 9 of the 13 fractions, their regression yields an intercept corresponding to an age $4564.42 \pm 0.12$ Myr, and MSWD $= 1.18$ (assuming ${}^{238}{\rm U}/{}^{235}{\rm U} = 137.88$), 
{\Steve 
%implying an intercept 
%$r = 0.6237875 \pm 0.000052$.
%}
We concur with their selection of points and confirm their regression, finding $4564.40 \pm 0.13$ Myr, with MSWD $= 1.14$.
}
The ${}^{238}{\rm U}/{}^{235}{\rm U}$ ratio in {\Steve whole-rock samples of} D'Orbigny has been measured by \citet{BrenneckaWadhwa2012}, who found $137.780 \pm 0.021$, and \citet{TissotEtal2017}, who found $137.793 \pm 0.025$.
We take the weighted mean of these, or $137.785 \pm 0.016$, and use this to update the age and its uncertainty {\Steve according to equations 3 and 4}.
%Every decrease of $0.001$ in the ${}^{238}{\rm U}/{}^{235}{\rm U}$ ratio corresponds to a younger age, by $0.0105$ Myr, so this new ${}^{238}{\rm U}/{}^{235}{\rm U}$ ratio decreases the age by 1.00 Myr, and introduces a new source of uncertainty of $\pm 0.17$ Myr that must be added in quadrature to the $\pm 0.12$ Myr above.
%We find \mbox{\boldmath$4563.42 \pm 0.21 \, {\rm Myr}$}. 
We find $4563.42 \pm 0.21$ Myr.

{\Steve
An important caveat is that the ${}^{238}{\rm U}/{}^{235}{\rm U}$ ratio was measured in whole-rock samples, whereas the Pb-Pb ages were determined using data primarily from Pb in (clino)pyroxene (especially the points with the most radiogenic Pb).
This matters because the U in pyroxene appears to be isotopically light, most probably because of isotopic fractionation during magmatic differentiation; \citet{TissotEtal2017} model this process and conclude that pyroxene grains have $\delta^{238}{\rm U}$ roughly $0.25\permil$ below the value for $\delta^{238}{\rm U}$ in the silicate melt in both D'Orbigny and Angra dos Reis.
Whether the ${}^{238}{\rm U}/{}^{235}{\rm U}$ ratio in the pyroxene grains that figure into the isochron is well represented by the ratio in the bulk samples depends on $f_{\rm cpx}$, the fraction of all U residing in clinopyroxene, with 
\begin{equation}
\delta^{238}{\rm U}_{\rm cpx} = \delta^{238}{\rm U}_{\rm bulk} - (1 -f_{\rm cpx}) \, (0.25\permil).
\end{equation}
\citet{TissotEtal2017} estimate $f_{\rm cpx} \approx 50\%$ in most angrites and similar achondrites, so that using whole-rock U isotopic ratios overestimates the ages of these achondrites, by about $0.19$ Myr, but possibly by up to $0.38$ Myr ($f_{\rm cpx} = 0$), or not at all ($f_{\rm cpx} = 1$).
We therefore favor 
\mbox{\boldmath$4563.24 \pm 0.21 \, {\rm Myr}$}.
}

\subsubsection{\it SAH 99555 (quenched angrite)}
SAH 99555 is a quenched angrite with an unshocked, fine-grained texture composed of anorthite, Al-Ti-bearing hedenbergite, olivine and mm-sized vesicles \citep{Keil2012}.

We take the value $(\alratio)_0 = (3.64 \pm 0.18) \times 10^{-6}$ determined by \citet{SchillerEtal2015}. 
This yields a time of formation \mbox{\boldmath$\Delta t_{26} = 5.14 \pm 0.05 \, {\rm Myr}$}.

The Pb isotopic ratios of SAH 99555 were determined by both \citet{Amelin2008b} and \citet{ConnellyEtal2008}.
\citet{Amelin2008b} built an isochron using 8 out of 10 whole-rock fractions, rejecting two clear outliers.
Assuming ${}^{238}{\rm U}/{}^{235}{\rm U} = 137.88$, they found an age $4564.86 \pm 0.38$ Myr, MSWD $= 1.5$. 
{\Steve
We essentially reproduce this regression, finding $4564.88 \pm 0.10$ Myr, MSWD $= 1.5$.}
% , and intercept $r = 0.623995 \pm 0.000043$.
\citet{ConnellyEtal2008} built an isochron using 8 of 11 whole-rock leachates, plus the residue from the pyroxene samples. 
Assuming ${}^{238}{\rm U}/{}^{235}{\rm U} = 137.88$, they found an age $4564.58 \pm 0.14$ Myr, MSWD $= 0.99$. 
{\Steve
We again essentially reproduce this regression, finding $4564.58 \pm 0.07$ Myr, MSWD $= 1.17$.
%, and intercept $r = 0.623865 \pm 0.000030$.
The weighted mean of the ages is $4564.614 \pm 0.131$ Myr.
%, implying an intercept $r = 0.623871 \pm 0.000057$. 
}
For the U isotopic rate, we use
the weighted mean of the values measured by \citet{TissotEtal2017}, 
${}^{238}{\rm U}/{}^{235}{\rm U} = 137.805 \pm 0.029$, and by \citet{ConnellyEtal2012}, 
${}^{238}{\rm U}/{}^{235}{\rm U} = 137.784 \pm 0.024$ (after {\Steve renormalizing to the same U standard as} \citealt{TissotEtal2017}), and adopt ${}^{238}{\rm U}/{}^{235}{\rm U} = 137.793 \pm 0.019$.
{\Steve
With this correction, we find an age {\mbox{$4563.70 \pm 0.24 \, {\rm Myr}$}.}
}

{\Steve 
Again we note the caveat that the ${}^{238}{\rm U}/{}^{235}{\rm U}$ ratio is measured in the bulk sample, not the pyroxene grains, meaning the value of $t_{\rm Pb}$ should be reduced by about 0.19 Myr.
We therefore favor 
\mbox{\boldmath$4563.51 \pm 0.24 \, {\rm Myr}$}.
}

\subsubsection{\it NWA 1670 (quenched angrite)}
NWA 1670 is a quenched angrite with a porphyritic texture including large olivine megacrysts in a fine-grained matrix of olivine, pyroxene, kirsch-steinite and anorthite as well as other accessory minerals \citep{Keil2012}.
{\Steve These indicate rapid cooling at $\sim 300 \, {\rm K} \, {\rm hr}^{-1}$ \citep{MikouchiEtal2003}.
}

It was analyzed by \citet{SchillerEtal2015}, who found  $(\alratio)_0 = (5.92 \pm 0.59) \times 10^{-7}$, which yields a time of formation \mbox{\boldmath$\Delta t_{26} = 4.63 \pm 0.10 \, {\rm Myr}$}.

\citet{SchillerEtal2015} also measured Pb isotopic ratios to determine a Pb-Pb age of $4565.39 \pm 0.24$ Myr, assuming ${}^{238}{\rm U}/{}^{235}{\rm U} = 137.786 \pm 0.013$.
{\EDIT We note that this value appears to have been taken from whole-rock measurements of other angrites \citep{ConnellyEtal2012}.
Besides this,}
{\Steve we were not able to confirm this age, because it is unclear which points to include in the regression.}
This isochron was based on using only six of the 13 available fractions, with no clear criteria for including or excluding points from the regression.
As a result, the uncertainty in the age has been considerably underestimated, as we now demonstrate. 

As described above, the absolute age of a sample can be found by performing a linear regression on the data $y_i = {}^{207}{\rm Pb}/{}^{206}{\rm Pb}$ vs. $x_i = {}^{204}{\rm Pb}/{}^{206}{\rm Pb}$, where each point $i$ represents an isotopic measurement in a different wash, leachate, or residue of a sample after it has been dissolved in various acids. 
Provided there is only one non-radiogenic source of lead---either primordial lead or terrestrial contamination---the data should array along a line, and the $y$-intercept of that line yields the absolute age according to Equation~\ref{eq:inverseisochron}.
Primordial lead is a likely component of meteorites and inclusions, and terrestrial contamination is pervasive 
\citep{GopelEtal1994,PalkEtal2018}, so it is likely that one of the fractions may contain an anomalous amount of one source of Pb or the other, and fall off the isochron.
{\Steve It is almost always necessary to exclude one or more points from the Pb-Pb isochron regression, and this is indeed the case for all the other Pb-Pb isochrons discussed in this paper.}
For example, in their analyses of D'Orbigny and SAH 99555, \citet{Amelin2008a} and \citet{Amelin2008b}, excluded from their regressions data points with elevated common Pb, or low total Pb.
In other analyses of Asuka 881394 \citep{WadhwaEtal2009,WimpennyEtal2019}, NWA 2976 \citep{BouvierEtal2011b}, and NWA 6704 \citep{AmelinEtal2019}, points were excluded essentially only if the ${}^{206}{\rm Pb}/{}^{204}{\rm Pb}$ was below a threshold value (indicating too little radiogenic Pb).
%Finally, for their CAI, \citet{BouvierWadhwa2010} regressed all the last leachates and residues (L7 and R)

In contrast, \citet{ConnellyEtal2008} (SAH 99555), \citet{SchillerEtal2015} (NWA 1670), and \citet{BollardEtal2017} (NWA 5697 chondrules) did not include or exclude points based on stated criteria such as dissolution order or threshold ${}^{206}{\rm Pb}/{}^{204}{\rm Pb}$.
We believe they instead excluded up to half their data based on whether the data fell off a pre-determined isochron.
This procedure is prone to confirmation bias, as we illustrate using the example of NWA 1670.
%ED: I think you should be clear that 'we believe' that they excluded data based on whether it was on the line. Unless they say this in the paper? You can probably come up with something better than 'we believe'
% SD, You're right Emilie, they just didn't say, and that's just how we interpret things.

\begin{center}
\begin{figure}[ht!]
\includegraphics[width=0.99\textwidth,angle=0]{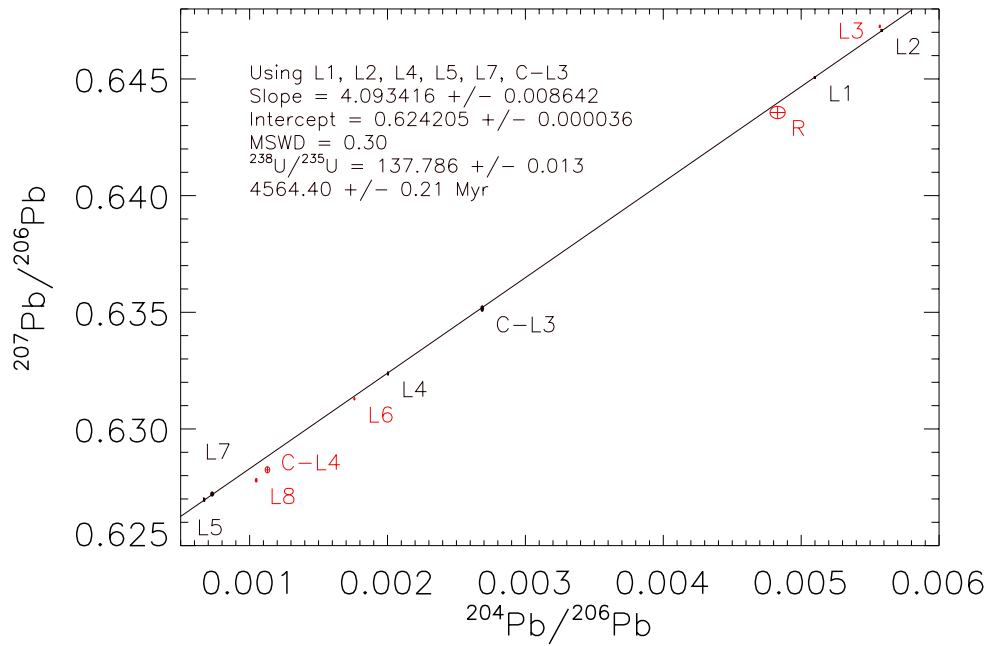}
\caption{Pb-Pb isochron of NWA 1670, based on the six data points regressed by \citet{SchillerEtal2015}, in black, with excluded points in red. The intercept of this isochron, and its uncertainty, yield an age $4564.40 \pm 0.21$ Myr.
\label{fig:nwa1670orig}}
\end{figure}
\end{center}

\citet{SchillerEtal2015} obtained Pb isotopic data for NWA 1670 from the following fractions: two acid washes of whole-rock samples (W1, W2), eight leachates (L1-L8) and a residue (R) of whole-rock samples, and two leachates (C-L3, C-L4) of clinopyroxene grains.
In {\bf Figure~\ref{fig:nwa1670orig}} we plot these 13 data (except for W1 and W2, with very high ${}^{204}{\rm Pb}/{}^{206}{\rm Pb} \approx 0.01$).
When we perform a York regression \citep{YorkEtal2004} 
%ED do you need to cite York here?)
% SD: That's a really good question, Emilie! I'm going to err on the side of doing that. I'm actually kind of tired of people saying they did the regression using the "Isoplot Model 1" (which is the York regression) technique and not explaining what that means. Frankly, I'd like to get everybody to understand they are using the York regression and be suspicious of anyone who doesn't.
on the same 6 data points selected by \citet{SchillerEtal2015} (L1, L2, L4, L5, L7, C-L3), we infer a slope %$4.093416 \pm 0.008642$
$4.093 \pm 0.009$ and intercept $0.624205 \pm 0.000038$, with MSWD $= 0.30$. 
%ED: do you need that many significant figures? Would 4.093±0.009 be enough?
% SD: You're right, Emilie.
With ${}^{238}{\rm U}/{}^{235}{\rm U} = 137.786 \pm 0.013$, we infer a Pb-Pb age of $4564.40 \pm 0.22$ Myr. 
This is almost identical to the result obtained by \citet{SchillerEtal2015}, who found an age $4564.39 \pm 0.24$ Myr, with MSWD $=0.31$. 

\begin{center}
\begin{figure}[ht!]
\includegraphics[width=0.99\textwidth,angle=0]{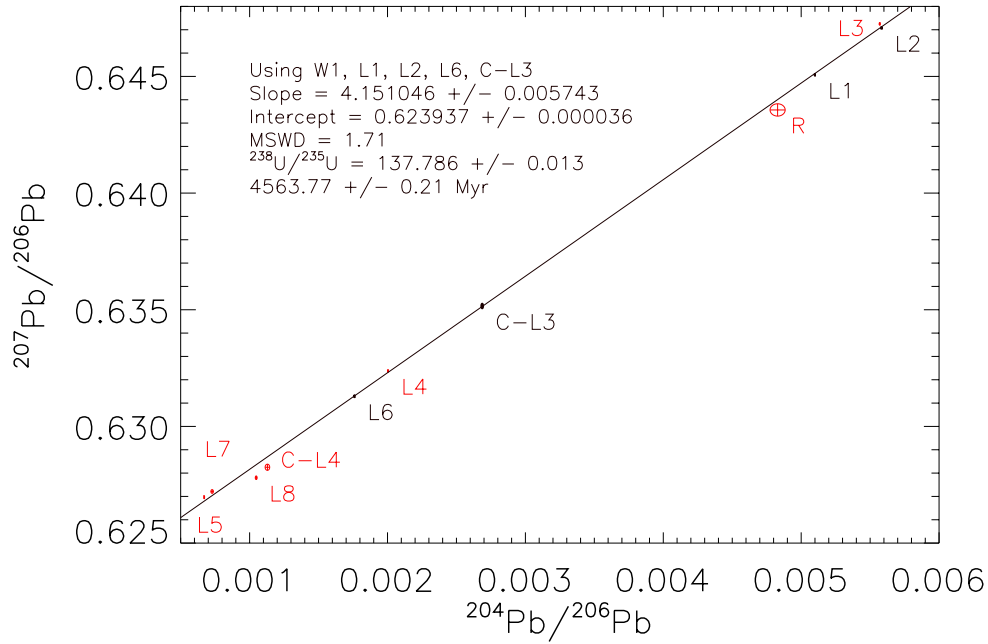}
\caption{Pb-Pb isochron of NWA 1670, based a different subset of data points. The intercept of this isochron, and its uncertainty, yield an age $4563.77 \pm 0.21$ Myr, 0.63 Myr younger than the age from Figure 1.
\label{fig:nwa1670low}}
\end{figure}
\end{center}

Except perhaps for W1 and W2, the reasons for excluding data points from the regression were not based on wash order or ${}^{204}{\rm Pb}/{}^{206}{\rm Pb}$ ratio, as this would not explain why L4 and L7 were included, but L6 and L8 were not, or why C-L3 was included but C-L4 was not. 
\citet{SchillerEtal2015} state that the reasons for excluding data points was the need to have components that only sampled radiogenic and terrestrial Pb, not initial Pb. 
They point out that the regression line is consistent with mixing between radiogenic and terrestrial Pb; we also find that this regression line lies only 0.0038 above the terrestrial Pb point (${}^{204}{\rm Pb}/{}^{206}{\rm Pb} = 0.0542$, ${}^{207}{\rm Pb}/{}^{206}{\rm Pb} = 0.84228$; \citealt{StaceyKramers1975}).
\citet{SchillerEtal2015} interpret points above the line (e.g., L3) to contain minor amounts of initial Pb, and points below the line (e.g., L6, L8, R, C-L4) to have lost radiogenic Pb. 
Notably, these interpretations are not based on any physical characteristics of the samples, but based on where they lie relative to the regression line.
Thus, this interpretation is circular logic. Points were excluded from the regression; those points that were not included in the regression fell off the regression line; for that reason alone, their exclusion was justified.
The fits obtained by such analyses have low MSWD, but significantly, an MSWD of only 0.30 is ``too good to be true".
The probability of six data points with random measurement errors exhibiting a fit with such a low MSWD is $\approx 5\%$ \citep{WendtCarl1991}, which is usually the threshold for acceptability.
This entire approach is prone to confirmation bias. 
%---------------------------------
% Figure 1
%ED comment: Steve is it possible to make the data points and data labels bigger? And maybe make the axis labels smaller font, the numbers on the 204/206 axis are pretty squished. And same with the next figure. If you want, I can try making this figure in matlab to see how it looks. 
% SD: Good suggestions, Emilie. I tweaked it and it looks better. 
%---------------------------------
Indeed, we have found a range of combinations of points that yield more probable isochrons (in terms of MSWD) and a wide variety of ages.
In {\bf Figure~\ref{fig:nwa1670low}}, we show a regression using a different set of five points (W1, L1, L2, L6, C-L3). 
We infer a slope %$4.151046 \pm 0.005743$
$4.151 \pm 0.006$ and intercept $0.623937 \pm 0.000036$, with MSWD $= 1.71$. 
%ED again I think too many sig figs
This line passes only 0.006 above terrestrial Pb, and a regression of five points with MSWD $=1.71$ is actually {\EDIT just as} likely as one with 0.30. 
The points that lie below the line (L8, R, C-L4) can be equally interpreted as having lost radiogenic Pb, and the points above the line (L3, L4, L5, L7) as having incorporated initial Pb. 
Thus, this is an equally valid isochron, but the age derived from it is $4563.77 \pm 0.21$ Myr, 0.63 Myr younger than the age derived in Figure 1.
%ED: could you add to this last sentence: "but the age derived from it is 4563 Myr, which is # Myr earlier than with the other data points."
% SD: Good idea.

%---------------------------------
% Figure 2
%---------------------------------

\begin{center}
\begin{figure}[ht!]
\includegraphics[width=0.99\textwidth,angle=0]{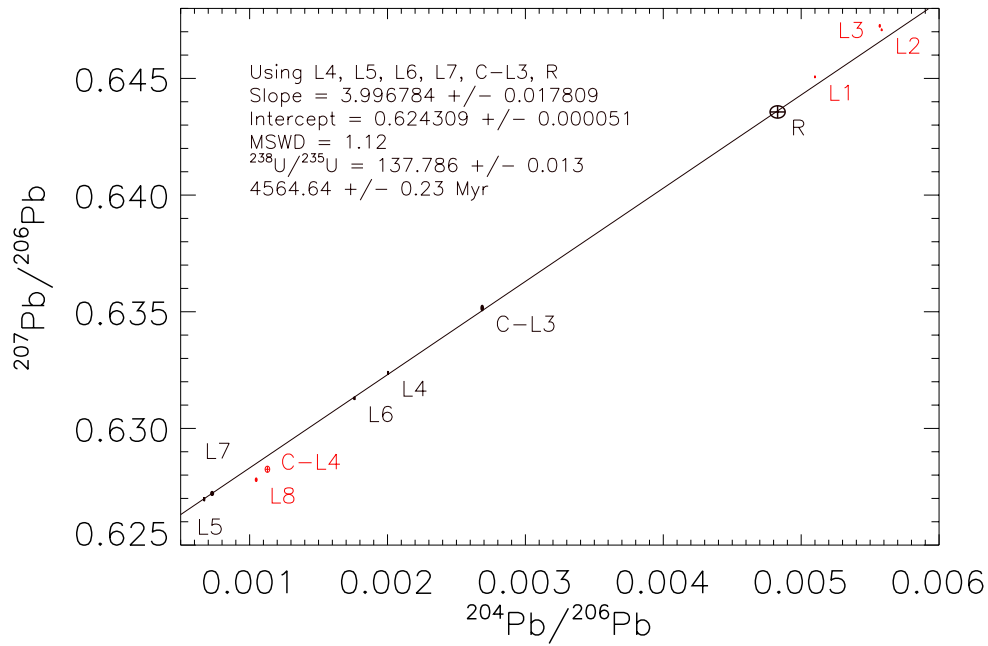}
\caption{Pb-Pb isochron of NWA 1670, based a different subset of data points. The intercept of this isochron, and its uncertainty, yield an age $4564.64 \pm 0.23$ Myr, 0.24 Myr older than the age from Figure 1.
\label{fig:nwa1670high}}
\end{figure}
\end{center}

Likewise, in {\bf Figure~\ref{fig:nwa1670high}}, we show a regression using a different set of six points (L4, L5, L6, L7, R, C-L3). 
We infer a slope %$3.996784 \pm 0.017809$
$4.000 \pm 0.018$ and intercept $0.624309 \pm 0.000051$, with MSWD $= 1.12$. 
This line passes only 0.001 below terrestrial Pb, and a regression of six points with MSWD $=1.12$ is very probable.
The points that lie below the line (L8, C-L4) can be equally interpreted as having lost radiogenic Pb, and the points above the line (L1, L2, L3) as having incorporated initial Pb. 
This isochron is as valid as the others, perhaps even more so because it passes closest to terrestrial Pb, has the {\EDIT most probable} MSWD, and there is some order to which fractions are included in the regression. 
The age derived from this regression is $4564.64 \pm 0.23$ Myr, 0.24 Myr older than the isochron derived in Figure 1.
%ED: do you really need to do two examples (Fig. 2 and 3) of using Schiller's data to prove that they excluded points without logic? I think that one example is good enough
% SD: Oh, the point is that one gives a much older age, the other gives a much younger age!  It's not piling on, it's important to establish the range!  But if that was lost, then I should try to explain it better.

%---------------------------------
% Figure 3
%---------------------------------

Through these examples, we have established that two isochrons equally valid to the one derived by \citet{SchillerEtal2015} yield a range of ages, from 0.63 Myr younger to 0.24 Myr older.
These isochrons are equally valid, because they use equal numbers of apparently equally valid points. 
Without objective criteria like ${}^{204}{\rm Pb}/{}^{206}{\rm Pb}$ ratio or Pb concentration to select points, 
points included in our regressions can equally well be argued to be valid because they lie on the regression line; points excluded from the regression can equally well be argued to exhibit loss of radiogenic Pb or excess of primordial Pb.
Without further information, we consider the 95\% confidence interval of possible Pb-Pb ages of NWA 1670 to extend from 4563.55 to 4564.87 Myr, and we take its Pb-Pb age to be \mbox{$4564.21 \pm 0.66 \, {\rm Myr}$}.
This encompasses the age $4564.39 \pm 0.24$ Myr reported by \citet{SchillerEtal2015}, but acknowledges a larger uncertainty.
{\Steve This age uncertainty may be magnified, depending on whether the isochron draws mostly from pyroxenes or whole rock washes and leachates, because of isotopic fractionation of U isotopes.}
%ED: to me, this argument gets a little lengthy and like you keep poking the bear over and over again. Can you remove the paragraph starting with "in the regression" or at least tone it down? 
% SD: OK, Emilie, fair enough. I've tried to improve it. I wonder what you think. I do think it's important to explain what's going on, very thoroughly, because it happens for the Bollard chondrules and the Connelly CAIs as well. This is a useful example. 

{\Steve 
Again we note the caveat that the ${}^{238}{\rm U}/{}^{235}{\rm U}$ ratio is measured in the bulk sample, not the pyroxene grains, meaning the value of $t_{\rm Pb}$ should be reduced by about 0.19 Myr.
We therefore favor 
\mbox{\boldmath$4564.02 \pm 0.66 \, {\rm Myr}$}.
}

\subsubsection{Asuka 881394 (eucrite-like achondrite)}
Asuka 881394 is a eucrite-like achondrite with a coarse-grained igneous texture with near equal amounts of anorthite and pyroxene. {\DRD The granoblastic texture suggests post-formation, low-grade metamorphism could have affected Asuka 881394 \citep{WimpennyEtal2019}.} While originally classified as a cumulate eucrite \citep{Takeda1997}, Asuka 881394 has geochemcial and isotopic qualities that preclude classification as such. As discusssed by \citet{WimpennyEtal2019}, the major element chemistry of the primary phases do not resemble cumulate eucrites (e.g., its plagioclase is too calcic, and the Mg-rich pyroxenes do not show evidence of inversion textures). Additionally, the $\Delta^{17}{\rm O}$ oxygen isotope composition of Asuka 881394 is 15${\sigma}$ above the mean value for HEDs \citep{ScottEtal2009}. 
% Cool info!

The Al-Mg systematics of Asuka 881394 have been measured by three groups: \citet{NyquistEtal2003}, who found $(\alratio)_0 = (1.18 \pm 0.14) \times 10^{-6}$, reanalyzed by \citet{WimpennyEtal2019} as $(1.18 \pm 0.31) \times 10^{-6}$; \citet{WadhwaEtal2009}, who found $(\alratio)_0 = (1.28 \pm 0.07) \times 10^{-6}$, eliminating one outlier from the regression; and \citet{WimpennyEtal2019}, who found $(\alratio)_0 = (1.48 \pm 0.12) \times 10^{-6}$.
We take the weighted average of these, $(\alratio)_0 = (1.31 \pm 0.06) \times 10^{-6}$. % $(\alratio)_0 = (1.307 \pm 0.056) \times 10^{-6}$. 
This yields a time of formation \mbox{\boldmath$\Delta t_{26} = 3.82 \pm 0.04 \, {\rm Myr}$}.

Pb-Pb dating of Asuka 881394 was done by \citet{WimpennyEtal2019}, who built an isochron using their pyroxene and whole-rock residue data, plus one whole-rock wash point, plus the most radiogenic residues (${}^{206}{\rm Pb}/{}^{204}{\rm Pb} > 400$) from the analysis by \citet{WadhwaEtal2009}.
They also determined ${}^{238}{\rm U}/{}^{235}{\rm U} = 137.786 \pm 0.038$ {\Steve in the bulk rock}, and found a Pb-Pb age of \mbox{$4564.95 \pm 0.53 \, {\rm Myr}$}, based on an isochron with MSWD $= 1.4$.
% I don't think we tried to reproduce this one.

{\Steve 
Again we note the caveat that the ${}^{238}{\rm U}/{}^{235}{\rm U}$ ratio is measured in the bulk sample, not the pyroxene grains, meaning the value of $t_{\rm Pb}$ should be reduced by about 0.19 Myr.
We therefore favor 
\mbox{\boldmath$4564.76 \pm 0.53 \, {\rm Myr}$}.
}

\subsection{NWA 7325 (ungrouped achondrite)}
NWA 7325 is an ungrouped achondrite with a medium-grained cumulate texture consisting of Mg-rich olivine, Cr-bearing diopside and Ca-rich plagioclase \citep{GoodrichEtal2017}.
The positive Eu anomaly in NWA 7325 is a geochemical sign of being a cumulate rock, but also can be interpreted as the result of melting or of a basaltic or gabbroic lithology \citep{BarratEtal2015}.

The Al-Mg systematics of NWA 7325 were measured by \citet{KoefoedEtal2016}, who found $(\alratio)_0 = (3.03 \pm 0.14) \times 10^{-7}$, which yields a time of formation \mbox{\boldmath$\Delta t_{26} = 5.33 \pm 0.05 \, {\rm Myr}$}.

\citet{KoefoedEtal2016} also analyzed its Pb-Pb systematics, restricting their regression to pyroxene residues, and building an isochron that excluded those points with ${}^{206}{\rm Pb}/{}^{204}{\rm Pb} < 50$, as they included obvious terrestrial contamination.
They did not measure the U isotopic ratio, instead adopting a value ${}^{238}{\rm U}/{}^{235}{\rm U} = 137.794$ as representative of materials from the inner Solar System \citep{GoldmannEtal2015}.
\citet{KoefoedEtal2016} found a Pb-Pb age for NWA 7325 of \mbox{\boldmath$4563.4 \pm 2.6 \, {\rm Myr}$}.

{\Steve More recently, \citet{CartwrightEtal2016metsoc} analyzed NWA 8486, which is paired with NWA 7325. 
Combining the data for the two achondrites, and apparently assuming the same U isotopic ratio, they found an age of $4563.9 \pm 1.7$ Myr.
As it is consistent with the \citet{KoefoedEtal2016} result but provides a more extreme test, we adopt this value with its smaller uncertainties.}

{\Steve Again we note the caveat that the ${}^{238}{\rm U}/{}^{235}{\rm U}$ ratio is measured in the bulk sample, not the pyroxene grains, meaning the value of $t_{\rm Pb}$ should be reduced by about 0.19 Myr. We therefore favor \mbox{\boldmath$4563.7 \pm 1.7 \, {\rm Myr}$}}.

\subsubsection{NWA 2976 (ungrouped carbonaceous achondrite)}

NWA 2976 {\Steve (paired with NWA 011)} is an unshocked, unbrecciated, ungrouped achondrite with coarse-grained pigeonite surrounded by fine-grained plagioclase with prevalent 120$^{\circ}$ triple junctions. \citep{YamaguchiEtal2002}.
The $^{54}$Cr, $^{50}$Ti and ${\Delta}^{17}$O isotope systematics of {\Steve NWA 011 have strong affinities to carbonaceous chondrites, and to the CR clan of chondrites in particular \citep{FlossEtal2005,SanbornEtal2019}.
The major element data presented by \citet{FlossEtal2005} point to oxidizing conditions during formation of NWA 011/2976.}

The Al-Mg systematics of NWA 2976 were measured by \citet{BouvierEtal2011b}, who found $(\alratio)_0 = (3.94 \pm 0.16) \times 10^{-7}$.
\citet{SchillerEtal2010} also reported $(\alratio)_0 = (4.91 \pm 0.46) \times 10^{-7}$ in NWA 2976.
\citet{SugiuraYamaguchi2007} reported  $(\alratio)_0 = (6.93 \pm 2.12) \times 10^{-7}$ in the paired achondrite NWA 011.
{\Steve We adopt the weighted mean of the values from \citet{BouvierEtal2011b} and \citet{SchillerEtal2010}, $(4.05 \pm 0.15) \times 10^{-7}$ and find a time of formation \mbox{\boldmath$\Delta t_{26} = 5.03 \pm 0.04 \, {\rm Myr}$}. 
}

For NWA 2976, \citet{BouvierEtal2011b} regressed the five whole-rock and pyroxene samples obtained from the last dissolution step, with ${}^{206}{\rm Pb}/{}^{204}{\rm Pb} > 1010$.
They measured ${}^{238}{\rm U}/{}^{235}{\rm U} = 137.751 \pm 0.018$, and  found a Pb-Pb age for NWA 2976 of $4262.89 \pm 0.59$ Myr from an isochron with MSWD $=0.02$.
{\Steve 
We agree with their choice of data to regress and reproduce their regression.
%, with intercept $r = 0.62372 \pm 0.00024$.
%We were able to reproduce their regression, finding an intercept $r = 0.6237218 \pm 0.000058$, which would yield an age $4562.91 \pm 0.23$ Myr for their U isotopic ratio, with MSWD $=0.02$.
%We cannot account for the large uncertainty found by \citet{BouvierEtal2011b}.
\citet{ConnellyEtal2012} measured an isotopic ratio 
${}^{238}{\rm U}/{}^{235}{\rm U} = 137.787 \pm 0.011$.
We combine these, finding ${}^{238}{\rm U}/{}^{235}{\rm U} = 137.777 \pm 0.009$, which then yields an age 
\mbox{\boldmath$4563.16 \pm 0.57 \, {\rm Myr}$}, 
}

{\Steve
As before, it is important to note that the U isotopic ratio was measured in whole-rock samples, and may not represent the isotopic ratio in the pyroxenes.
Here we argue, though, that for the two `carbonaceous' achondrites NWA 2976 and NWA 6704, likely $f_{\rm cpx} \approx 1$, most of the U probably resides in the pyroxene grains, and the whole-rock isotopic ratio probably is appropriate.
Both of these achondrites are more oxidized. 
As members of the CC isotopic reservoir, they would be expected to contain more water and other volatiles. 
Both \citet{WarrenEtal2013} and \citet{HibiyaEtal2019} estimated that the CC achondrite NWA 6693, paired with NWA 6704, formed under oxygen fugacity conditions 2 log units above the iron-w\"{u}stite buffer ($\Delta {\rm IW} = +2$), as opposed to angrites, which formed at $\Delta {\rm IW} = +1$ \citep{BrettEtal1977,JurewiczEtal1993,McKayEtal1994, TissotEtal2022}.
More importantly, \citet{IzawaEtal2022lpsc} concluded that the paired achondrite NWA 011, {\EDIT paired with NWA 2976}, experienced hydrous magmatism and a water-bearing melt. 

Under these redox conditions, U is expected to be mostly in the U({\sc iv}) $4^{+}$ valence state (as insoluble UO$_2$) with some in the U({\sc vi}) $6^{+}$ state as the soluble ${\rm UO}_{2}^{2+}$ uranyl ion.  %\citep{SchreiberAnders1980,Halse2014}
\citep{ChevreuxEtal2021}.
At IW+2, compared to IW+1, slightly more U speciates as U({\sc vi}) instead of U({\sc iv}), but the more important factor is the presence of H$_2$O in the magma. 
This could drastically change the partitioning of U due to the ability of water to depolymerize the aluminosilicate network and produce Non-Bridging Oxygen (NBO) sites.
These are the coordination sites by which U (of either valence) can enter a growing crystal. 
We expect that in a hydrous magma, U will readily partition into the major crystallizing phases, as opposed to remaining in the melt to partition into late-forming phosphates as in an anhydrous magma.
As a result, the pyroxene grains will take up most the U and share the same U isotopic ratio as the whole rock. 
Assuming $f_{\rm cpx} \approx 1$, we do not apply an age correction to NWA 2976 or NWA 6704.
By our reasoning, it should not be applied to any achondrite formed from hydrous magmatism.
}

\subsubsection{NWA 6704 (ungrouped carbonaceous achondrite)}

NWA 6704 is an unshocked, ungrouped achondrite with a medium-grained, cumulate texture comprised of low-Ca pyroxene along with Ni-rich olivine and sodic plagioclase \citep{HibiyaEtal2019}. {\DRD \citet{WarrenEtal2013} argue the coarse texture and bulk subchondritic MgO/SiO$_2$ indicate origin as an igneous cumulate.
However, \citet{HibiyaEtal2019} argue that these same textural features are best explained by a rapid initial crystallization, which agrees with the apparent absence of geochemical evidence for significant magmatic differentiation. 
The Ni-rich phases along with the V/(Al+Cr) ratio in the spinels suggest formation at relatively oxidizing oxygen fugacity $\Delta{\rm IW} = +2$, 2 log units above the Iron-W\"{u}stite) buffer \citep{WarrenEtal2013}. 
Both \citet{HibiyaEtal2019} and \citet{WarrenEtal2013} demonstrate the volatile-depleted nature of NWA 6704/6693. 
\citet{SanbornEtal2019} showed the ${\Delta}^{17}$O, ${\epsilon}^{50}$Ti and ${\epsilon}^{54}$Cr isotope composition of NWA 6704/6693 agree with the CR clan of chondrites.}
%The $^{54}$Cr, $^{50}$Ti and ${\Delta}^{17}$O isotope systematics have strong affinities to carbonaceous chondrites, specifically CR-type chondrites \citep{SanbornEtal2019,HibiyaEtal2019}.

The Al-Mg systematics of NWA 6704 were measured by \citet{SanbornEtal2019}, who found $(\alratio)_0 = (3.15 \pm 0.38) \times 10^{-7}$, the value we adopt, which yields a time of formation \mbox{\boldmath$\Delta t_{26} = 5.29 \pm 0.13 \, {\rm Myr}$}.

The Pb-Pb age of NWA 6704 was determined by \citet{AmelinEtal2019}, who regressed 13 pyroxene fraction data with ${}^{206}{\rm Pb}/{}^{204}{\rm Pb} > 700$, excluding one outlier.
Based on a measured ${}^{238}{\rm U}/{}^{235}{\rm U} = 137.7784 \pm 0.0097$, their regression yielded an age \mbox{\boldmath$4562.76^{+0.22}_{-0.30} \, {\rm Myr}$}.

{\Steve As with NWA 2976, we assume the U isotopic ratio in the pyroxenes is well represented by the whole rock ratio, and do not apply a correction.}

\subsection{NWA 5697 (L3) Chondrules}

The Al-Mg and Pb-Pb systematics were simultaneously measured by \citet{BollardEtal2017} and \citet{BollardEtal2019} for eight chondrules from the carbonaceous chondrite Allende (CV3) and the ordinary chondrite NWA 5697 (L3).
However, although \citet{BollardEtal2017} reported Pb-Pb dates for eight chondrules analyzed for Al-Mg, only four of these (NWA 5697: {\it 2-C1}, {\it 3-C5}, {\it 5-C2}, {\it 11-C1}) were U-corrected using a ${}^{238}{\rm U}/{}^{235}{\rm U}$ measured in the chondrule; the rest (NWA 5697: {\it 5-C10}, {\it C1}, {\it C3}; and Allende: {\it C30}) assumed a bulk chondrite value ${}^{238}{\rm U}/{}^{235}{\rm U} = 137.786 \pm 0.013$.
Because some chondrules are resolvably different from this assumed value (e.g., NWA 5697 {\it 5-C2} has ${}^{238}{\rm U}/{}^{235}{\rm U} = 137.756 \pm 0.029$), we do not assume all chondrules should have this value, and we restrict our analysis only to those four measured ordinary chondrite chondrules. 
As well, \citet{BollardEtal2017} constructed their isochrons using a subset of the fractions.
As with the case of NWA 1670, we have found it necessary to re-analyze the Pb-Pb isochrons of these four chondrules. 
Each of these recalculated Pb-Pb ages of the chondrules encompasses the values reported by \citet{BollardEtal2017}, but acknowledges a much greater uncertainty. 

{\it Chondrule 2-C1}:
For this chondrule, \citet{BollardEtal2017} regressed 11 out of 20 fractions (L3, L4, L6, L8, L9, R, W11-2, L3-2, L4-2, L7-2, L8-2) and reported an age $4567.57 \pm 0.56$ Myr and MSWD $= 1.20$.
Performing the same regression, we find a similar $4567.45 \pm 0.48$ Myr and MSWD $= 1.12$.
Regressing a different subset of 11 fractions (L3, L4, L5, L6, L8, L9, R, L4-2, L7-2, L8-2, L9-2) yields an age $4567.33 \pm 0.48$ Myr and MSWD $= 1.88$.
And regressing yet another subset of 11 fractions (L1, L2, L7, L9, R, L2-2, L3-2, L4-2, L7-2, L8-2, L9-2) yields an age $4567.85 \pm 0.50$ Myr and MSWD $= 0.44$.
We take the age of chondrule 2-C1 to be \mbox{\boldmath$4567.60 \pm 0.75 \, {\rm Myr}$}.
\citet{BollardEtal2019} found $(\alratio)_0 = (7.56 \pm 0.13) \times 10^{-6}$ (MSWD=1.3), which yields a time of formation \mbox{\boldmath$\Delta t_{26} = 2.00 \pm 0.21 \, {\rm Myr}$}. 

{\it Chondrule 3-C5}:
For this chondrule, \citet{BollardEtal2017} regressed 10 out of 14 fractions (W11, L1, L2, L3, L4, L5, L6, L7, L8, L9) and reported an age $4566.20 \pm 0.63$ Myr and MSWD $= 1.27$.
Performing the same regression, we find a similar $4566.13 \pm 0.51$ Myr and MSWD $= 1.29$.
Simply removing point L8 from the regression, we find an age $4565.54 \pm 0.54$ Myr and MSWD $= 0.46$.
Regressing instead a different set of nine points (W11, L2, L4, L5, L6, L7, L8, L11, l12), we find instead an age $4566.57 \pm 0.56$ Myr and MSWD $= 0.63$.
We take the age of chondrule 3-C5 to be \mbox{\boldmath$4566.07 \pm 1.07 \, {\rm Myr}$}.
\citet{BollardEtal2019} found $(\alratio)_0 = (7.04 \pm 1.51) \times 10^{-6}$ (MSWD$=0.9$, which yields a time of formation \mbox{\boldmath$\Delta t_{26} = 1.84 \pm 0.21 \, {\rm Myr}$}. 

{\it Chondrule 5-C2}:
For this chondrule, \citet{BollardEtal2017} regressed 8 out of 15 fractions (W11, L3, L4, L6, L7, L8, L9, L10) and reported an age $4567.54 \pm 0.52$ and MSWD $=0.66$, a result we reproduce exactly.
Regressing a subset of 9 fractions (L1, L2, L3, L4, L6, L7, L9, L10, L11) yields an age $4566.84 \pm 0.56$ Myr and a more probable MSWD $= 0.94$. It is not clear why the other points should be rejected. 
Regressing a subset of eight data points (W10, W11, L2, L3, L4, L6, L8, L10, L11) yields an age $4567.70 \pm 0.49$ Myr and MSWD $= 0.72$. 
We take the age of chondrule 5-C2 to be 
\mbox{\boldmath$4567.24 \pm 0.95 \, {\rm Myr}$}. %\mbox{\boldmath$4567.24 \pm 0.99 \, {\rm Myr}$}. 
\citet{BollardEtal2019} found $(\alratio)_0 = (8.85 \pm 1.83) \times 10^{-6}$ (MSWD$=1.8$), which yields a time of formation after $t\!\!=\!\!0$ of \mbox{\boldmath$\Delta t_{26} = 2.07 \pm 0.22 \, {\rm Myr}$}. 

{\it Chondrule 11-C1}:
Finally, for this chondrule, \citet{BollardEtal2017} regressed 8 out of 15 fractions (W11, L4, L5, L6, L7, L8, L9, L10) and reported an age $4565.84 \pm 0.72$ and MSWD $=1.16$.
We find similar but different $4565.64 \pm 0.55$ Myr, MSWD $= 1.29$.
Regressing a subset of eight data points (W11, L2, L4, L5, L7, L9, L10, L12) yields an age $4565.36 \pm 0.59$ Myr and MSWD $= 0.85$. 
Regressing a different subset of 10 fractions (W11, L2, L4, L5, L6, L7, L8, L9, L10, L12) yields an age $4565.69 \pm 0.56$ Myr and MSWD $= 1.77$. 
We take the age of chondrule 11-C1 to be \mbox{\boldmath$4565.51 \pm 0.74 \, {\rm Myr}$}. 
%\mbox{\boldmath$4565.50 \pm 0.72 \, {\rm Myr}$}. 
\citet{BollardEtal2019} found $(\alratio)_0 = (5.55 \pm 1.84) \times 10^{-6}$ (MSWD$=0.7$), which yields a time of formation \mbox{\boldmath$\Delta t_{26} = 2.32 \pm 0.34 \, {\rm Myr}$}. 

\subsection{CAIs}
\label{sec:caiisochrons}

{\Steve 
Although the age of $t\!\!=\!\!0$ is commonly taken to be the Pb-Pb age of CAIs, the assumption that the Pb-Pb chronometer closed at the same time that the Al-Mg did (at $t\!\!=\!\!0$) is untested.
Given that CAIs almost certainly spent millions of years in the same region of the solar nebula where chondrules formed, it is highly probable they could have been subject to transient heating events like those that formed chondrules.
As discussed above (\S 2.2), this could have reset the Pb-Pb chronometer without resetting the Al-Mg chronometer.}
With this in mind, we ask whether any of the CAIs from which \citet{ConnellyEtal2012} determined a Pb-Pb age (\textit{22E}, \textit{31E}, \textit{33E}, \textit{SJ101}), actually show evidence that the Pb-Pb chronometer closed at $t\!\!=\!\!0$, when $(\alratio)_0 \approx 5 \times 10^{-5}$.

To our knowledge, Al-Mg systematics have not been measured in CAI \textit{33E}.
\citet{LarsenEtal2011} measured bulk isotopic ratios in CAIs \textit{22E} and \textit{31E}, but not an internal isochron; if a late resetting of {\Steve the Al-Mg system in} these CAIs occurred, this would not have been detected by bulk measurements.

This leaves only CAI \textit{SJ101}.
{\Steve \cite{AmelinEtal2010} measured ${}^{238}{\rm U}/{}^{235}{\rm U} = 137.876 \pm 0.043$ and an isochron with intercept $0.625000 \pm 0.000092$ (MSWD $=1.07$) and derived a Pb-Pb age of $4567.18 \pm 0.50$ Myr for this CAI.}
Its Al-Mg systematics were measured by \citet{MacPhersonEtal2017}, who found $(\alratio)_0 = (5.2 \pm 0.5) \times 10^{-5}$. 
{\Steve It is widely assumed that the event recorded by the Pb-Pb chronometer is the same as the one recorded by the Al-Mg chronometer, i.e., the CAI's formation.
This in turn rests on the assumption that any thermal event resetting the Pb-Pb chronometer would cause diffusion of Mg and reset the Al-Mg chronometer, as well. 
As discussed in \S 2.2, however, a typical chondrule-forming transient heating event would have reset the Pb-Pb chronometer, and also the Al-Mg system in anorthite, but not in melilite, pyroxene, or spinel.  
It is highly significant that the isochron built by \citet{MacPhersonEtal2017} is built from these three minerals, and that it clearly shows that anorthite is slightly disturbed, with points in anorthite falling below the isochron.
The CAI \textit{SJ101} data admit the possibility that it formed and its Al-Mg chronometer was set at $t\!\!=\!\!0$, but that it was later transiently heated to peak temperatures $\approx 1750$ K and cooling rates $\approx 500 \, {\rm K} \, {\rm hr}^{-1}$ like chondrules.
It cannot be known that the Pb-Pb system closed at the same time as the Al-Mg system.
}

{\Steve For completeness, we also examine the Pb-Pb isochrons of other CAIs.}
For the CAI \textit{22E}, \citet{ConnellyEtal2012} reported a Pb-Pb age of $4567.35 \pm 0.28$ Myr, based on  a regression with MSWD $= 0.91$ and measured $\uratio = 137.627 \pm 0.022$. 
They generated 20 different leachates/washes/residues in their analysis, but based the intercept on a regression involving only 11 of these points (L1, L3, L6, L7, L8, L9, L10, L11, W7, W8, W9). 
Performing a York regression on these data points, and propagating the uncertainties in the $\uratio$ ratio and intercept in quadrature, we find a similar age $4567.32 \pm 0.27$ Myr, on an isochron with MSWD $= 1.08$. 
As was the case for NWA 1670, we were able to find many combinations of 11 data points that yielded acceptable isochrons (MSWD $< 1.94$, the maximum for 11 points; \citealt{WendtCarl1991}).
For one set of 11 points (W1, W5, W6, W7, W8, W9, L3, L6, L9, L10, L11), we found MSWD $= 0.29$, and an age $4567.21 \pm 0.29$ Myr. 
For a different set of 11 points (W1, W5, W6, W7, W9, L1, L3, L4, L6, L7, L8), we found MSWD $= 0.77$ and an age $4567.54 \pm 0.31$ Myr. 
As with the case of NWA 1670, no physical criteria were laid out for the exclusion of data points.
\citet{ConnellyEtal2008} argued that the points they included in the regression eliminated are free of contamination by terrestrial Pb, on the basis that the regression line extends to primordial Pb, and they eliminated points below this line (i.e., toward terrestrial Pb). 
%Third, the vertex of their triangle does not quite lie at the position of
%primordial Pb, despite their claim that ``This isochron projects back to an
%initial Pb isotopic composition that [is] slightly less radiogenic than the
%initial Pb isotopic composition of the solar system as defined by troilite 
%from the 1AB [IAB] iron meteorite Nantan."
%Primordial Pb lies at 
%${}^{204}{\rm Pb}/{}^{206}{\rm Pb} = 0.107459 \pm 0.000016$, 
%${}^{207}{\rm Pb}/{}^{206}{\rm Pb} = 1.10759 \pm 0.00010$ 
%\citep{BlichertToftEtal2010}.
%When we regress their data using the 11 points they selected, we find an 
%isochron with slope $4.461878 \pm 0.002292$ and intercept 
%$0.626185 \pm 0.000235$. 
%Every isochron within those uncertainties falls below the primordial Pb 
%point, with a small but definite minimum distance to it of 
%$(4.23 \pm 0.41) \times 10^{-4}$. 
However, their regression line falls below primordial Pb, suggesting {\EDIT (by their criteria)} that actually all their data have some terrestrial contamination. 
Moreover, some included data (e.g., W7, L7) in fact plot above the regression line. 
In the absence of independent criteria for judging the isochrons, we consider all the isochrons described above to be equally valid, and conclude that the 95\% confidence interval for the Pb-Pb age of CAI \textit{22E} should extend from 4566.92 to 4567.85 Myr, i.e., $4567.39 \pm 0.47$ Myr. 
We find basically the same Pb-Pb age for \textit{22E} as \citet{ConnellyEtal2012}, but find the uncertainty has been underestimated by about a factor of almost 2. 
We find similar outcomes for CAIs \textit{31E} and \textit{32E}.

We find no issue with the reported Pb-Pb ages of the other CAIs \textit{SJ101}, \textit{B1}, or \textit{B4}.
{\Steve It is worth noting that CAI \textit{B1} records $(\alratio)_0 = (5.03 \pm 0.26) \times 10^{-5}$ \citep{BouvierWadhwa2010}.
}
We conclude that CAIs actually do record Pb-Pb ages over a range of times from about 4568.2 Myr to 4567.2 Myr, 
{\Steve and that their Pb-Pb ages may have been reset about 1 Myr after they formed.}

\subsection{Summary}

In {\bf Table~\ref{table:adoptedages}} we convert the $(\alratio)_0$ initial ratios of each sample into a time of formation after $t\!\!=\!\!0$ assuming an abundance $(\alratio)_{\rm SS} = 5.23 \times 10^{-5}$ at $t\!\!=\!\!0$ and mean-life 1.034 Myr, and we compile our adopted Pb-Pb ages, $t_{\rm Pb}$, for each achondrite and chondrule, as well as the implied value of $t_{\rm SS} = t_{\rm Pb} + \Delta t_{26}$.
{\Steve For the first five achondrites we have applied the 0.19 Myr correction of \citet{TissotEtal2017}, making the Pb-Pb ages younger.
We have not applied this to the achondrites NWA 2976 or NWA 6704, because we assume their whole-rock U isotopic ratios reflect the composition of the pyroxene grains in which the Pb-Pb ages were determined.
We have not applied the correction to chondrules because ${}^{238}{\rm U}/{}^{235}{\rm U}$ was measured in them directly.
}
The uncertainties in $t_{\rm SS}$ are found by adding the uncertainties in $t_{\rm Pb}$ and $\Delta t_{26}$ in quadrature.

%-------------------------------------------
% TABLE 1. Adopted Ages
%$\!\!\!\!\!\!\!\!\!\!\!\!\!\!\!\!\!\!\!\!\!\!\!\!\!\!\!\!\!\!\!\!$
\begin{table}[h!]
\noindent
\begin{minipage}{4.25in}
\footnotesize
\caption{Adopted $\Delta t_{26}$ and Pb-Pb ages  ages of seven bulk achondrites and four chondrules, plus the implied value of $t_{\rm SS} = t_{\rm Pb} + \Delta t_{26}$. \vspace{0.1in}
\label{table:adoptedages}}

\begin{tabular}{l|c|c|c}
{Sample} & $\Delta t_{26}$ (Myr) & $t_{\rm Pb}$ (Myr) & $t_{\rm SS}$ (Myr) \\
\hline
{D'Orbigny} & 
$5.06 \pm 0.10$ & $4563.24 \pm 0.21$ &
$4568.30 \pm 0.23$ \\
{SAH 99555} &
$5.14 \pm 0.05$ & $4563.51 \pm 0.24$ & 
$4568.65 \pm 0.25$ \\
{NWA 1670} &
$4.64 \pm 0.10$ & $4564.02 \pm 0.66$ &
$4568.66 \pm 0.67$ \\
{Asuka 881394} & 
$3.82 \pm 0.04$ & $4564.76 \pm 0.53$ &
$4568.58 \pm 0.53$ \\
{NWA 7325} & 
$5.33 \pm 0.05$ & $4563.71 \pm 1.7$ &
$4569.04 \pm 1.7$ \\
\hline
{NWA 2796} & 
$5.03 \pm 0.04$ & $4563.17 \pm 0.57$ &
$4568.20 \pm 0.57$ \\
{NWA 6704} & 
$5.29 \pm 0.13$ & $4562.76 \pm 0.26$ &
$4568.05 \pm 0.29$ \\
\hline 
{NWA 5697 2-C1} & 
$2.00 \pm 0.21$ & $4567.60 \pm 0.75$ &
$4569.60 \pm 0.78$ \\
{NWA 5697 3-C5} & 
$1.84 \pm 0.21$ & $4566.07 \pm 1.07$ &
$4567.91 \pm 1.09$ \\
{NWA 5976 5-C2} & 
$2.07 \pm 0.22$ & $4567.24 \pm 0.95$ &
$4569.31 \pm 0.97$ \\
{NWA 5697 11-C1} & 
$2.32 \pm 0.34$ & $4565.51 \pm 0.74$ &
$4567.83 \pm 0.81$ \\
\hline
\end{tabular}

%\footnotesize{}
\end{minipage}
\end{table}
%-------------------------------------------

{\Steve These samples all imply values of $t_{\rm SS}$ ranging from 4567.8 and 4569.6 Myr, centered around 4568.7 Myr.
The average uncertainty in the Pb-Pb model ages $t_{\rm SS}$ is 0.8 Myr, due almost entirely to the uncertainties in the Pb-Pb ages of each sample.}

\section{Determination of $t_{\rm SS}$}

{\Steve
\subsection{Statistical analysis}
}

As seen from Table~\ref{table:adoptedages}, the majority of achondrites and chondrules imply that $t\!\!=\!\!0$ of the Solar System occurred more than 4568 Myr ago. 
The best estimate of the Pb-Pb age of $t\!\!=\!\!0$, $t_{\rm SS}^{*}$, is the weighted average of the estimates of $t_{{\rm SS},i}$ from each sample (indexed by $i$):
\begin{equation}
t_{\rm SS}^{*} = \left[ \sum_{i=1}^{N} w_i \right]^{-1} \, \times \,
 \sum_{i=1}^{N} \, w_i \left[ t_{{\rm Pb},i} + \tau_{26} \, \ln \left( R_{26,{\rm SS}} / R_{26,i} \right) \right],
\end{equation}
where $R_{26,i}$ and $t_{{\rm Pb},i}$ are the initial ratio $(\alratio)_0$ and Pb-Pb age in sample $i$, $w_i = 1 / \sigma_{i}^2$, 
{\Steve $\sigma_{t26,i}^2 = \tau_{26}^{2} ( \sigma_{R26,i} / R_{26,i})^2$},
and $\sigma_{i}^{2} = \sigma_{t26,i}^2 +\sigma_{t{\rm Pb},i}^{2}$ accounts for the uncertainties in the ages.
The uncertainty in the weighted mean is $\left[ \sum_{i=1}^{N} w_{i} \right]^{-1/2}$.
{\Steve Although the uncertainties in each sample's model age are typically 0.8 Myr, by averaging several estimates of $t_{\rm SS}$ together, the uncertainty in $t_{\rm SS}$ can be reduced.
This treatment gives the best estimate (and the uncertainty in our estimate) of $t_{\rm SS}^{*}$.
}

{\Steve 
Whether the ages actually are concordant with that estimate of $t_{\rm SS}{^*}$ must be determined by calculating the $z$ scores of each age.
We first determine for each sample $i$ our best estimate of when that achondrite or chondrule formed, $\Delta t_{i}$, assuming both the Al-Mg and Pb-Pb systems closed simultaneously.
We calculate $\Delta t_{{\rm Pb},i} = t_{\rm SS}^{*} - t_{{\rm Pb},i}$ and 
\begin{equation}
\Delta t_{i} = \left( \frac{1}{\sigma_{t26,i}^2} + \frac{1}{\sigma_{t{\rm Pb},i}^2} \right)^{-1} \, \left(
\frac{\Delta t_{26,i}}{\sigma_{t26,i}^2} + \frac{\Delta t_{{\rm Pb},i}}{\sigma_{t{\rm Pb},i}^2} 
\right).
\end{equation}}
{\Steve The $z$ scores measure the departure of the measured time of formation of a sample from this modeled time of formation, in units of $1\sigma$:
\begin{equation}
z_{26,i} = 2 \frac{\Delta t_{26,i} - \Delta t_{i}}{\sigma_{t26,i}}
\end{equation}
and
\begin{equation}
z_{{\rm Pb},i} = 2 \frac{\Delta t_{{\rm Pb},i} - \Delta t_{i}}{\sigma_{t{\rm Pb},i}},
\end{equation}
where the factor of 2 recognizes that the $\sigma$ are reported as $2\sigma$ errors.
If $\left|z_{26,i}\right| > 2$ or $\left|z_{{\rm Pb},i}\right| > 2$, then it is improbable ($< 5\%$) that that single time of formation $\Delta t_{26,i}$ or $\Delta t_{{\rm Pb},i}$ is being adequately fit by the data.
}

{\Steve Of course, in a dataset with $> 20$ ages being fit, it would be improbable {\it not} to have at least one age with $\left| z \right| > 2$. 
This is accounted for by calculating the $\chi_{\nu}^2$ or MSWD quantity for the dataset:
\begin{equation}
\chi_{\nu}^{2} = \frac{1}{2 A - 1} \, \sum_{i=1}^{A} \left( z_{26,i}^2 + z_{{\rm Pb},i}^2 \right),
\end{equation}
where $A$ is the number of samples (achondrites, chondrules, etc.) being considered, and it is recognized that the number of data being fit is $N = 2 A$ (two ages for each sample), and the number of degrees of freedom in the dataset is 1 (only $t_{\rm SS}^{*}$ is being fit).
This quantity must not exceed a critical value $\chi_{\nu,{\rm max}}^2 \approx 1 + 2 (2 / (N-1))^{1/2}$ or else there is a $< 5\%$ probability that the dataset as a whole is being adequately fit by the single value of $t_{\rm SS}^{*}$ \citep{WendtCarl1991}.
}

{\Steve
\subsection{Achondrites}
}

{\Steve
We first consider just the seven achondrites as a group, since we are relatively certain that each achondrite cooled rapidly enough that the Al-Mg and Pb-Pb systems closed simultaneously and were not reset.
We find $t_{\rm SS}^{*} = 4568.39$ Myr.
Among this set of 14 times of formation to be fit, we find that two (the Pb-Pb ages of SAH 99555 and NWA 6704) are barely fit, at the $2.0\sigma$ and $2.1\sigma$ levels; then again, in a sample of 14 formation times, we would expect 5\%, or about 1, to be $> 2.0\sigma$ discrepant.
Overall, the data are fit very well, with $\chi_{\nu}^2 = 0.98 < \chi_{\nu,{\rm max}}^2 = 1.72$ (48\% probability).
These data are concordant.
This is our preferred case.
}

{\Steve
The above analysis assumes a correction of 0.19 Myr applied to Pb-Pb ages of the five achondrites that did not involve hydrous silicate melts, as suggested by \citet{TissotEtal2017} for angrites, assuming $f_{\rm cpx} = 0.5$ in NC achondrites and $f_{\rm cpx} = 1$ in CC achondrites.
Had we assumed $f_{\rm cpx} = 0$ in the NC achondrites and applied a correction of -0.38 Myr, the fit would be improved: $t_{\rm SS}^{*} = 4568.24$ Myr, with all 14 $z$ scores $< 2$, and $\chisq = 0.57$.
Had we assumed $f_{\rm cpx} = 1$ in the NC achondrites and applied no correction, the fit would be worsened: $t_{\rm SS}^{*} = 4568.52$ Myr, with the Pb-Pb ages of SAH 99555 and NWA 6704 discordant at the $> 2.5\sigma$ and $> 2.9\sigma$ levels, and $\chisq = 1.63$ (7\% probability).
If $f_{\rm cpx} < 1$ for the CC achondrites, this would significantly worsen the fits.
We do not treat them as free parameters, but we consider it likely that $f_{\rm cpx} \approx 0.5$ {\it at most} in the NC achondrites and $f_{\rm cpx} \approx 1$ in the CC achondrites, due to the hydrous nature of the melt.
We adopt these values of $f_{\rm cpx}$ and age corrections going forward. 
}

{\Steve
\subsection{Chondrules}
}
{\Steve
We next consider just the four chondrules. 
We fit these eight formation times with $t_{\rm SS}^{*} = 4568.76$ Myr, but barely.
Two of the Pb-Pb ages have $z > 2.0$, and $\chisq = 1.94$, barely below the critical value 2.07, indicating a probability $\approx 6\%$ that the chondrule data are concordant with each other.
However, the two values of $t_{\rm SS}^{*}$ implied by the achondrites (4568.39 Myr) and chondrules (4568.76 Myr) are only 0.34 Myr apart, suggesting the achondrites and chondrules might be concordant with each other.
}

{\Steve
If we adopt the value of $t_{\rm SS}^{*} = 4568.39$ Myr from the achondrites fit (or 4568.36 Myr from Paper II), the four chondrules are fit well, except for one anomalous $z$ score, the Pb-Pb formation time of \textit{2-C1} being discordant at the $3.0\sigma$ level. 
We find $\chisq = 1.35 < \chisq_{\rm max} = 1.56$ (13\% probability).
% While technically concordant, the poor fit of \textit{2-C1} suggests the chondrules are not consistent with this value of $t_{\rm SS}^{*}$.
}

{\Steve
If we average the seven achondrites and four chondrules together, we find $t_{\rm SS}^{*} = 4568.42$ Myr.
Among the 14 achondrite formation times, only one (the Pb-Pb age of NWA 6704) is off, at the $2.3\sigma$ level.
Among the eight chondrule formation times, only the Pb-Pb age of \textit{2-C1} is off, at the $2.9\sigma$ level.
That is, two out of 22 formation times have $z$ scores $> 2$, which is close the value (1.1) expected statistically. 
For this ensemble, $\chisq = 1.35$, which does not exceed $\chisq_{\rm max} = 1.56$ for these 22 data points, and is significant (probability 12\%).
This fit is slightly better for the chondrules, not much worse for the achondrites, and can be considered concordant, although the Pb-Pb age of \textit{2-C1} is a concern.
}

{\Steve
We can vary $t_{\rm SS}^{*}$ about this value and ask what range of $t_{\rm SS}^{*}$ yields $\chisq \leq 1.62$.
We find the range is 4568.31 to 4568.54 Myr, implying the Pb-Pb age of $t\!\!=\!\!0$ is roughly \mbox{\boldmath$4568.42 \pm 0.12 \, {\rm Myr}$}.
}

{\Steve
\subsection{CAIs}
}

{\Steve
For none of the values of $t_{\rm SS}^{*}$ above is the formation times of CAI \textit{SJ101} concordant with the others. 
This would require $t_{\rm SS}^{*} = 4567.17 \pm 0.51$ Myr, 1.3 Myr younger than our preferred value.
For our preferred $t_{\rm SS}^{*} = 4568.42$ Myr, the Pb-Pb $z$ score of \textit{SJ101} is discordant at the $5.2\sigma$ level.
}

{\Steve
\subsection{Effects of varying half-lives}
}

These calculations assume a half-life of $\altwosix$ of 0.717 Myr, but the value is not known this precisely.
As reviewed by \citet{Nishiizumi2003}, estimates had converged by the 1970s at $\approx 0.72$ Myr, but with some range. 
\citet{RightmireEtal1958} reported $0.738 \pm 0.29(1\sigma)$ Myr based on gamma ray measurements; but this was superseded by the work of \citet{SamworthEtal1972}, who revised the branching ratios and reported $0.716 \pm 0.032(1\sigma)$ Myr.
\citet{NorrisEtal1983} undertook independent gamma ray spectrometry and found $0.705 \pm 0.024(1\sigma)$ Myr. 
\citet{MiddletonEtal1983} used activity accelerator mass spectrometry to find 0.699 and 0.705 Myr using two different standards; they recommended $0.702 \pm 0.056(1\sigma)$ Myr.
Finally, \citet{ThomasEtal1984} found $0.78 \pm 0.05(1\sigma)$ Myr based on a technique involving the gamma ray cross section.
\citet{AuerEtal2009} took the weighted mean of these values to derive $0.717 \pm 0.017(1\sigma)$ Myr, and this is commonly cited.
The value listed in the chart of the nuclides \citep{WalkerEtal1989}, 0.73 Myr, also is commonly cited, but the source of this value is not clear. 
We adopt the \citet{AuerEtal2009} value.
{\Steve
This is also the value listed in the compilation of \citet{Kondev2021}.
}

We have repeated the calculations above, varying the half-life of $\altwosix$ across its allowed range $0.717 \pm 0.034 (2\sigma)$ Myr. 
For a half-life of 0.683 Myr, we find 
{\Steve
$t_{\rm SS}^{*} = 4568.17$ Myr,for the $A=7$ samples.
}
For a half-life of 0.751 Myr, we find 
{\Steve
$t_{\rm SS}^{*} = 4568.63$ Myr.
}
The $2\sigma$ uncertainty of $\pm 0.049$ Myr in the $\altwosix$ mean-life translates into a difference of {\Steve $\pm 0.22$} Myr in $t_{\rm SS}^{*}$.

{\Steve 
Of course the uncertainties in the ${}^{238}{\rm U}$ and ${}^{235}{\rm U}$ half-lives lead to $40\times$ larger uncertainties in the Pb-Pb times of formation, $\pm 9$ Myr \citep{TissotEtal2017}.
It is understood that the reported values of $t_{\rm SS}^{*}$ refer to the Pb-Pb age of a sample that closed at $t\!\!=\!\!0$, assuming standard half-lives (\S 1.1).
}
%We take $t_{\rm SS} = 4568.67 \pm 0.16$ Myr (or $\pm 0.29$ Myr including the uncertainty in the $\altwosix$ half-life) to be the Pb-Pb age of the Solar System, the value that would be obtained by dating samples, using commonly adopted uranium half-lives, if the Pb-Pb system in the samples closed at $t\!\!=\!\!0$. 
%---------------------------------
% Figure 5
\begin{center}
\begin{figure}[ht!]
$\!\!\!\!\!\!\!\!\!\!\!\!\!\!\!\!\!\!\!\!\!\!\!$
\includegraphics[width=0.99\textwidth,angle=0]{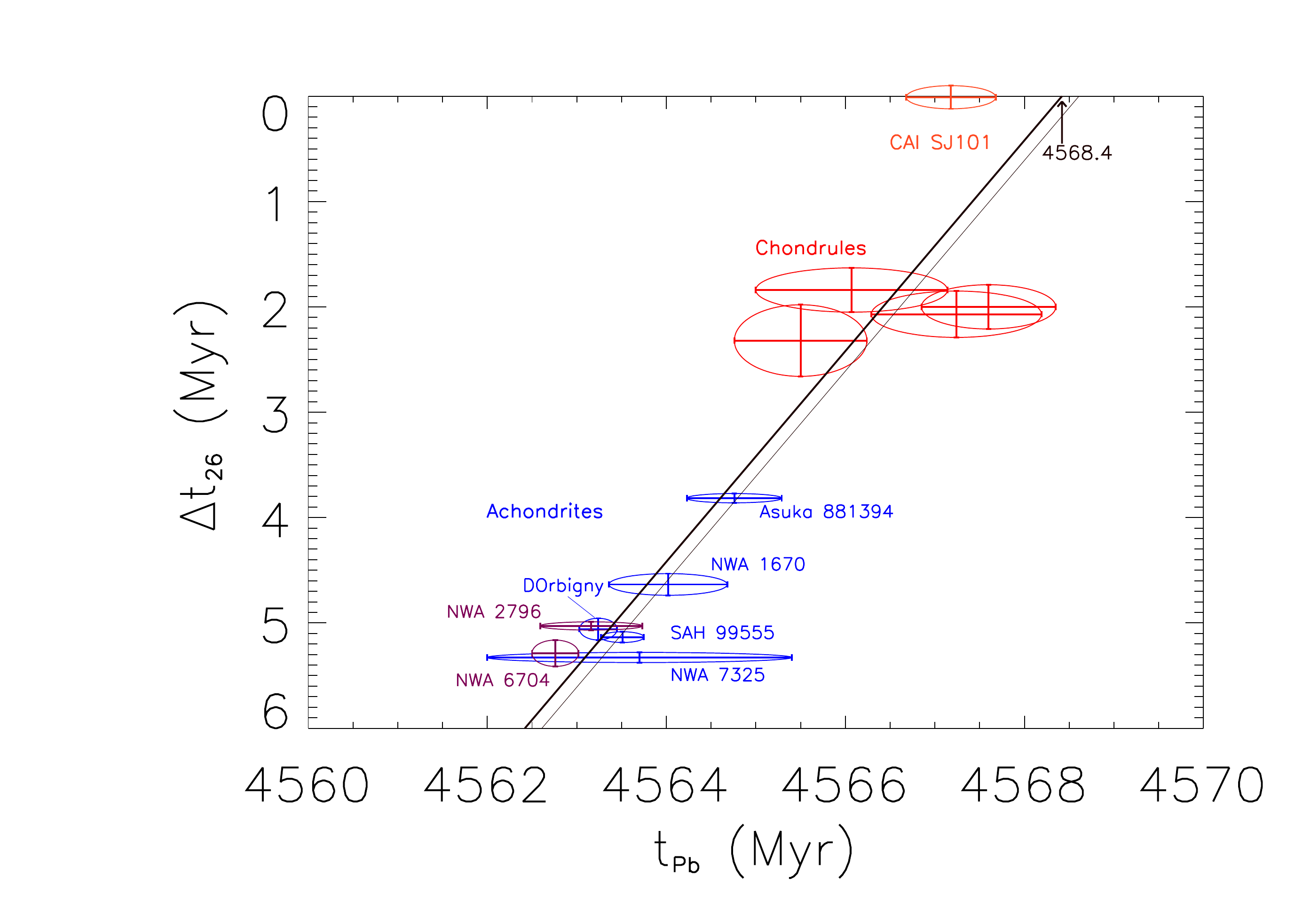}
\caption{Times of formation after $t\!\!=\!\!0$ implied by the Al-Mg system, $\Delta t_{26}$, and the age determined by the Pb-Pb system, $t_{\rm Pb}$, for the five NC achondrites (blue), two CC achondrites (purple), four chondrules (red), and one CAI (orange) in which both are measured. The black line denotes the locus of points yielding $t_{\rm Pb} + \Delta t_{26} = 4568.42$ Myr, the apparent value of $t_{\rm SS}$. 
All 14 formation times of the 7 achondrites are consistent with that value, except the Pb-Pb age of NWA 6704 is inconsistent at the $2.3\sigma$ level. 
{\EDIT The achondrite Erg Chech 002 has a Pb-Pb age \citep{RegerEtal2023} but two very different reported Al-Mg formation times \citep{BarratEtal2021,RegerEtal2023}; because of this discrepancy will be the subject of a forthcoming paper. 
}
The 8 chondrule formations times are marginally consistent with that value, except the Pb-Pb age of \textit{2-C1} is inconsistent at the $2.9\sigma$ level. Still, the overall goodness-of-fit parameter for the combined 22 ages is $\chisq = 1.36$, which is statistically significant (12\% probability). The ages are concordant.
The light black line is shifted to the right by 0.19 Myr, corresponding to $t_{\rm SS} = 4568.61$ Myr if the five NC achondrites' Pb-Pb ages had not been made 0.19 Myr younger to account for the isotopic fractionation of U in pyroxene grains relative to whole-rock measurements, as suggested by \citet{TissotEtal2017}. In that scenario, the ages would be barely concordant.
CAI \textit{SJ101} is not concordant with the other samples, which we attribute to a late resetting of the Pb-Pb chronometer without resetting the Al-Mg chronometer (\S 2.2).
\label{fig:tcai}}
\end{figure}
\end{center}
%---------------------------------

\section{Discussion}

{\Steve 
\subsection{Comparison to other treatments}
}

{\Steve 
Our approach is similar to, but distinct from, a few other attempts to determine the age of the Solar System's $t\!\!=\!\!0$ by statistically correlating Pb-Pb ages against the Al-Mg or other systems.}

\citet{LugmairShukolyukov1998} correlated Mn-Cr ages against Pb-Pb ages of several achondrites to determine $t_{\rm SS}$. 
They found a range of values 4568 to 4571 Myr, compared to the then-accepted age of CAIs, $4566 \pm 2$ Myr \citep{GopelEtal1991}, although again none of these Pb-Pb ages was U-corrected. 
According to this data, an age 4568 Myr would be consistent with both CAI formation and $t_{\rm SS}$.
{\Steve These Pb-Pb ages were not as precise as more modern methods allow, and were not in any event U-corrected.}

A different approach was taken by \citet{NyquistEtal2009} to estimate $t_{\rm SS}$ (their ``$T_{\rm SS}$").
They compiled $\log_{10}(\alratio)_0$ and $\log_{10}(\mnratio)_0$ ratios, as well as Pb-Pb ages, for several samples: the angrites D'Orbigny and Sahara (SAH) 99555; the eucrite-like Asuka 881394; ureilites DaG 165 and Dag 319; an estimate for the howardite-eucrite-diogenite (HED) parent body; and Semarkona chondrules.
It is difficult to convert $(\mnratio)_0$ ratios directly into $\Delta t_{53}$ ages because the initial abundance $(\mnratio)_{\rm SS}$ and even the half-life of $\mnfivethree$ are poorly known. 
\citet{NyquistEtal2009} did not correlate $\Delta t_{26}$ ages against Pb-Pb ages $t_{\rm Pb}$, but instead used the Al-Mg system to calibrate the Mn-Cr system, then found the Mn-Cr time of formation $\Delta t_{53}$ for one achondrite, Lewis Cliff (LEW) 86010, and used its measured Pb-Pb age $t_{\rm Pb}$ as an anchor, to derive $t_{\rm SS} = t_{\rm Pb} + \Delta t_{53}$. 

Specifically, \citet{NyquistEtal2009} regressed the Al-Mg and Mn-Cr data to show that the quantities were linearly correlated as 
\[
\log_{10}(\mnratio)_0 = (-5.485 \pm 0.028) 
\]
\begin{equation}
+ (0.23 \pm 0.04) \times \left[ \log_{10}(\alratio)_0 - (-6.252 \pm 0.074) \right]
\label{eq:nyquist}
\end{equation}
The slope in this linear correlation, $0.23 \pm 0.04(2\sigma)$, which should be the ratio of half-lives, \citet{NyquistEtal2009} estimated from measurements should be $0.20 \pm 0.02(1\sigma)$. % \citep{AuerEtal2009,HondaImamura1971}.
This suggests a $\mnfivethree$ half-life $3.12 \pm 0.65(2\sigma)$ Myr.
This linear correlation was extended to the value $(\alratio)_0 = 5.1 \times 10^{-5}$, the value they assigned to the solar nebula at $t\!\!=\!\!0$, to find the ratio in the solar nebula value then, $(\mnratio)_{\rm SS} = 9.14 \times 10^{-6}$. 
We note that this implicitly assumes that $\altwosix$ and $\mnfivethree$ were homogeneous (or identically heterogeneous) between the regions of CAI and achondrite formation. 
From there, using the inferred value $(\mnratio)_0 = 1.35 \times 10^{-6}$ in the achondrite LEW 86010, they calculated $\Delta t_{53} = 10.23$ Myr based on a $\mnfivethree$ half-life of 3.7 Myr, with 10\% uncertainty ($1\sigma$). 
Adopting a Pb-Pb age of $4558.55 \pm 0.15$ Myr for LEW 86010, \citet{NyquistEtal2009} then calculated $t_{\rm SS} = 4568.8 \pm 1.0$ Myr. 
\citet{NyquistEtal2009} repeated this calculation using a slope $0.20 \pm 0.02$ in the correlation above, finding $t_{\rm SS} = 4568.1 \pm 0.6$ Myr.
They then took the weighted mean of these two estimates to get their best estimate of $t_{\rm SS} = 4568.2 \pm 0.5$ Myr. 

It is notable that this value of $t_{\rm SS}$ is much older than the accepted Pb-Pb ages of CAIs measured today ($4567.30 \pm 0.16$ Myr), and even the majority of Pb-Pb ages of CAIs used in the analysis of \citet{NyquistEtal2009}.
However, there are many caveats that may have prevented adoption of this value as the age of the Solar System. 
First, of course, none of the Pb-Pb ages in their analysis was U-corrected, including those of LEW 86010 or the CAIs.
Thus the value of $t_{\rm SS}$ potentially could be shifted by about $> 1$ Myr \citep{BrenneckaWadhwa2012,TissotEtal2017}.
Second, the calculation of $t_{\rm SS}$ was done quite indirectly. 
Since $\altwosix$ was implicitly assumed to be homogeneous anyway, it might have been simpler and more precise to correlate $\Delta t_{26}$ against Pb-Pb ages directly, as \citet{NyquistEtal2009} did in their Figure 3; this, incidentally, would have yielded a value $t_{\rm SS} \approx 4569.3$ Myr.
Third, although some statistical approaches were employed, ultimately the value of $t_{\rm SS}$ relied on a single anchor, LEW 86010. 
This increases the uncertainty in $t_{\rm SS}$ due to the combined uncertainties in $(\mnratio)_{\rm SS}$, $\tau_{53}$, $(\mnratio)_0$ in LEW 86010, and the Pb-Pb age of LEW 86010. 
In fact, the $1\sigma$ uncertainty in the half-life of $\mnfivethree$ is $\pm 10\%$ \citep{HondaImamura1971}, and this dominates the uncertainty in $t_{\rm SS}$ in their treatment, so \citet{NyquistEtal2009} should have reported $t_{\rm SS} = 4568.2 \pm 1.0$ Myr. 

In similar fashion, \citet{SanbornEtal2019}, correlated times of formation derived from Al-Mg systematics, and Pb-Pb ages, for several achondrites and the CAIs listed above. 
They were focused on the slope (which yields a $\altwosix$ half-life of $0.69 \pm 0.12$ Myr), and did not report the intercept, necessary for finding the Pb-Pb age of $t\!\!=\!\!0$ when $(\alratio) = 5.23 \times 10^{-5}$. 
Nevertheless, it is clear by inspection of their Figure 6 that they would have inferred an age $4567.8$ Myr, with an uncertainty $\sim 0.5$ Myr, regressing the achondrite and CAI data.
Their regression would appear to be discordant or just barely concordant with the CAI Pb-Pb ages of \citet{ConnellyEtal2012}. 

{\Steve 
More recently, \citet{PirallaEtal2023} have attempted to determine $t_{\rm SS}$ using three approaches. 
In the first, they averaged the model ages of eight volcanic achondrites, ($t_{\rm Pb} + \Delta t_{26}$ or $t_{\rm Pb} + \Delta t_{182}$), like those reported in our Table 1. 
Through a Monte Carlo approach, they determined $t_{\rm SS} = 4568.50_{-0.99}^{+0.91} \, {\rm Myr}$ using the Al-Mg ages.
Their second approach was to compare Pb-Pb ages and Al-Mg formation times of their measured spinels and, essentially, anchor to their oldest chondrule. 
They found $t_{\rm SS} = 4568.36 \pm 0.59$ Myr. 
Their third approach was to combine a probabilistic approach to their use of the oldest chondrule, to derive $t_{\rm SS} = 4568.67_{-0.59}^{+0.79} \, {\rm Myr}$.
}

{\Steve
Our approach is similar in some ways to these methods, but is marked by many innovations. 
First, we have selected data based on the need to have measured, U-corrected Pb-Pb ages as well as Al-Mg ages from internal isochrons, for all samples; we abandon CAIs in favor of achondrites.
We have quantitatively demonstrated the possibility that the Al-Mg and Pb-Pb systems did not close simultaneously in CAIs that were transiently heated. 
We have carefully compiled and vetted the Pb-Pb isochrons to make sure they are being analyzed in the same fashion, and that they provide robust estimates of both the ages and the uncertainties. 
This involved applying examining how points were included or excluded from the isochron, properly averaging and propagating the errors when combining multiple analyses, etc.
We have applied the likely corrections of $\delta^{238}{\rm U} \approx -0.19$ Myr to the angrite-like achondrites, as advocated by \citet{TissotEtal2017}, where geochemically appropriate.
We have directly correlated Pb-Pb ages against Al-Mg ages, as others have, but {most importantly, we have  applied statistical techniques to assess the goodness of fit and concordancy}.
}

\subsection{Heterogeneity of ${}^{26}{\rm Al}$?}
 \label{sec:heterogeneity}
 
We have calculated an age of 
{\Steve $t_{\rm SS} = 4568.42 \pm 0.24$ Myr}
for the time at which the solar nebula was characterized by the canonical ratio $(\alratio) = 5.23 \times 10^{-5}$ (assuming homogeneity of $\altwosix$).
In contrast, the ages of CAIs \textit{E22}, \textit{E31}, \textit{SJ101}, and \textit{B4}, which are generally considered to record near-canonical $(\alratio)_0$ ratios \citep{LarsenEtal2011,MacPhersonEtal2017,BouvierWadhwa2010}, are $4567.3 \pm 0.2$ Myr \citep{ConnellyEtal2012} to $4568.2 \pm 0.2$ Myr \citep{BouvierEtal2011a}.
Because a homogeneous amount of $\altwosix$ at 4568.42 Myr would have decayed by a factor of {\Steve 3.3 by 4567.18 Myr}, it has been inferred that $\altwosix$ was not homogeneously distributed in the solar system \citep[e.g.,][]{LarsenEtal2011}. 

For example, \citet{BollardEtal2019} concluded that the reservoir from which achondrites formed was depleted in $\altwosix$ by a factor of {\Steve 3.8} compared to the CAI-forming region.
They reached this conclusion by obtaining the $(\alratio)_0$ in eight chondrules, taking the Pb-Pb ages $t_{\rm Pb}$ inferred for the same chondrules by \citet{BollardEtal2017}, then extrapolating back to the time of $t\!\!=\!\!0$ at $t_{\rm SS}$, to estimate the $\alratio$ of the material comprising each chondrule:
\begin{equation}
\left( \alratio \right)_{t=0} = (\alratio)_0 \, \exp \left( +( t_{\rm SS} - t_{\rm Pb}) / \tau_{26} \right).
\end{equation}
Taking $t_{\rm SS} = 4567.3$ Myr, they inferred values of
$(\alratio)_{t=0}$ ranging from $0.4 \times 10^{-5}$ to $2.7 \times 10^{-5}$.
The average was $1.36 \times 10^{-5}$, with hints of a bimodal distribution: the four lowest values averaged to $5 \times 10^{-6}$, and the four highest values to $1.8 \times 10^{-5}$.
% although they did not statistically test this, the chondrule ages did not seem compatible with a single average $(\alratio)_{t=0}$.
They concluded that $\altwosix$ was heterogeneous.

%However, the results presented here demonstrate that there is not strong evidence for heterogeneity of $\altwosix$, as the uncertainties in the Pb-Pb ages of the chondrules have undoubtedly been underestimated, sometimes by factors of 2 or more.
% Because of this, the various $(\alratio)_{t=0}$ implied by each chondrule tend to scatter within uncertainties around a single value.
% Moreover, that single value is compatible with the canonical value.
However, if one instead assumes {\Steve $t_{\rm SS} = 4568.42$ Myr}, then the CAIs must have taken {\Steve 1.24} Myr longer to form than previously thought, and their \linebreak $(\alratio)_{t=0}$ values should be multiplied by {\Steve $\exp(+1.24/1.034) = 3.3$}.
In that case, the average value of $(\alratio)_{t=0}$ implied by the eight chondrules would be {\Steve $4.5 \times 10^{-5}$, very close to} the canonical value. 

Conclusions of $\altwosix$ heterogeneity based on discordancy between the Al-Mg and Pb-Pb chronometers ultimately derive from the assumption that the Pb-Pb system closed at $t\!\!=\!\!0$, i.e., at the same time as the Al-Mg system. 
{\Steve
As we have demonstrated, the 22 Al-Mg and Pb-Pb ages of seven achondrites and four chondrules are made concordant in a statistical sense with a single value of $t_{\rm SS}$.
It could have been the case that no single value of $t_{\rm SS}$ could make the samples, especially the achondrites, concordant; this would have falsified the hypothesis of $\altwosix$ homogeneity.
Instead, the concordance of the data strongly suggest that $\altwosix$ was homogeneous and that CAI \textit{SJ101} and the other CAIs are the anomaly.
Given that transient heating events like those experienced by chondrules can reset the Pb-Pb chronometer without resetting the Al-Mg system, we consider CAIs to be unreliable testers of the homogeneity hypothesis.
}

{\Steve 
The consistency of the data with homogeneity of $\altwosix$ is exactly what is expected from astrophysical models, which do not generally predict conditions in which the CAI-forming region would have $3-4$ times the level of $\altwosix$ as other reservoirs.
}
Such large variations in $\alratio$ would not be expected in the molecular cloud, as molecular cloud cores take several Myr to collapse, during which time turbulent diffusion will readily mix materials across several parsecs \citep{PanEtal2012}, an expectation borne out by the surprising chemical similarity of stars born in the same cluster \citep{FengKrumholz2014,ArmillottaEtal2018}. 
On the scale of a molecular cloud core, $< 1$ pc across, there should be no differences in the composition of the accreting material.
Any spatial variations that did exist across a molecular cloud core would be further mixed during collapse \citep{KuffmeierEtal2017}.
Models have been proposed in which the infalling material varied over time; for example, to explain stable isotope anomalies, \citet{NanneEtal2019} proposed that early-accreted material might be richer in supernova-derived material than later-accreted material.
But in such models, heterogeneity in the molecular cloud is merely assumed {\Steve or asserted}, not demonstrated.
Moreover, as in the model of \citep{NanneEtal2019}, increased $\altwosix$ in the inner disk would require the cloud core interior to contain stellar ejecta, but not its exterior, an unlikely scenario.
% Because material closest to the cloud core center accretes first \citep{KuffmeierEtal2017}, this would demand the interior of a cloud core hold supernova-derived material but its exterior not, but how this situation would arise is not explained.

Because heterogeneities of $\altwosix$ cannot be inherited from the molecular cloud, they would have to result from late injections of $\altwosix$-bearing material from supernovas or Wolf-Rayet stars, into the protoplanetary disk.
Late injection has long been invoked as an explanation for why some hibonite-dominated grains apparently lacked $\altwosix$: presumably they formed before $\altwosix$ was injected \citep{SahijpalGoswami1998}.
However, the fact that {\it only} certain rare corundum- or hibonite-dominated inclusions exhibit a lack of $\altwosix$ (e.g., \citealt{KrotEtal2012}) more strongly suggests a chemical heterogeneity, as suggested by \citep{LarsenEtal2020}, and not spatial or temporal variations. 
As demonstrated by \citet{OuelletteEtal2007}, it is possible to inject dust grains from stellar ejecta into a protoplanetary disk; but it is highly improbable for a disk to form close enough to a stellar source to receive relevant amounts of ejecta \citep{OuelletteEtal2010}.
Although it has not been quantitatively explored whether CAIs could form with $4\times$ more $\altwosix$ than the rest of the disk, a simple argument suggests CAIs should form with less $\altwosix$, not more: over the scale of the disk, the stellar ejecta injects a uniform mass of $\altwosix$ per area; but that injected material is more diluted by greater surface densities in the inner disk where CAIs form. 

Approaching the problem from a different angle, the abundances of about a dozen short-lived radionuclides in the solar nebula are very successfully explained as being inherited from the molecular cloud (contaminated by supernovae and Wolf-Rayet winds), and not through late injection or irradiation in the solar nebula \citep{Young2020}.
This demands these isotopes, including $\altwosix$, would have been spatially well mixed in the solar nebula from before $t\!\!=\!\!0$.

Ascribing $\altwosix$ heterogeneities to something to do with star formation is too simplistic. 
Astrophysical models overall suggest $\altwosix$ should have been homogeneous in the Sun's protoplanetary disk, and that the discrepancy between the Al-Mg and Pb-Pb chronometers has to do with the Pb-Pb chronometer itself.
% SD: I started to realize that a few paragraphs like this are needed. They aren't intended to prove anything, but I think the non-astrophysicists need to hear some pushback from astrophysicists. 

\subsection{Relative vs.\ Absolute Ages}

The determination of {\Steve $t_{\rm SS} = 4568.42 \pm 0.24$ Myr} allows us to use the Pb-Pb data available for a number of achondrites, and unlock the potential of the Pb-Pb system as a {\it relative} chronometer. 
Relative ages are much more precise than absolute ages: absolute ages are uncertain to within $\pm 9$ Myr ($2\sigma$) due to the uncertainties in the half-lives of ${}^{235}{\rm U}$ and ${}^{238}{\rm U}$ \citep{Amelin2006,TissotEtal2017}.
These systematic uncertainties cancel when using the Pb-Pb system to calculate relative ages, and the precision can approach $\pm 0.3$ Myr, a factor of 30 more precise.
{\Steve Even ignoring the half-life uncertainties and comparing to the measurement uncertainties, the uncertainties in times of formation after $t\!\!=\!\!0$ are typically 0.13 Myr, 5 times better than the uncertainties in the model ages, which are typically 0.75 Myr.}  

Relative ages are also {\bf much} more relevant to models of the protoplanetary disk.
All astrophysical models of planet formation desperately need quantification of the order of events in the solar nebula, relative to a commonly accepted event at a defined $t\!\!=\!\!0$.
These could then be compared to astronomical constraints on the ages of young stellar objects, for which the timing of evolutionary stages is typically precise to within $< 1$ Myr \citep{HaischEtal2001,Mamajek2009}.
In contrast, there is almost no model of stellar or planetary evolution, that uses information of how long ago the protoplanetary disk stage was, that would be affected if it turned out the Solar System were 4560 Myr or 4580 Myr old instead of 4569 Myr old. 
For example, the tightest astrophysical constraints from helioseismology on the Sun’s age are $4600 \pm 40 (1\sigma)$ Myr \citep{HoudekGough2011}, and $4587 \pm 7$ Myr or $4569 \pm 6 (1\sigma)$ Myr, based on different databases of physical quantities \citep{BonannoFrohlich2015}.
For these allied fields of study, ``4.57 Gyr ago" is precise enough. 

%%We strongly advocate that dates of events in the solar nebula be reported as times relative to $t\!\!=\!\!0$ when the $\alratio$ in the solar nebula was a particular value; we take that value to be $5.23 \times 10^{-5}$.
%%We strongly advocate against the practice that is common in the meteoritics community, which is to first convert Al-Mg or other relative ages into absolute ages using an {\it anchor} meteorite, and then discern the order in which things occurred in the solar nebula using these absolute ages. 
%%Anchoring means finding the time of formation (after $t\!\!=\!\!0$) of an object like an achondrite, by using its experimentally determined ratio of, for example, $(\mnratio)_0$, to find its absolute age or time of formation relative to that of an anchor like the achondrite D’Orbigny, with its own measured $(\mnratio)_0$ ratio, then using the anchor's absolute Pb-Pb age to infer an absolute age of the achondrite.
% Of course, this absolute age is uncertain by up to $\pm 9$ Myr and only has significance as a time of formation relative to other objects, such as another achondrite, or CAIs.
%%Using the Pb-Pb age of $t\!\!=\!\!0$, the time of formation of the achondrite after $t\!\!=\!\!0$ could be found to be:
%%\begin{equation}
%%\Delta t_{53} = \tau_{53} \, \ln \left[ (\mnratio)_{\rm DOrbigny} / (\mnratio)_{0} \right]
%% + \left( t_{\rm SS} - t_{\rm Pb,DOrbigny} \right).
%% \label{eqn:anchor}
%%\end{equation}  
%%The practice of using anchors introduces considerable uncertainty on the way to figuring out the time at which events occurred.

{\Steve In addition, the need to choose anchors to report absolute ages introduces confusion and imprecision, especially to those not working in the meteoritics community, as a recent example from the literature illustrates.}
The two achondrites NWA 11119 and Erg Chech 002 are each among the most ancient crustal rocks in the meteoritic record. 
Which formed first? 
% Did they form at the same time? 
\citet{SrinivasanEtal2018} reported $(\alratio)_0 = (1.69 \pm 0.09) \times 10^{-6}$ in NWA 11119, while \citet{BarratEtal2021} reported $(\alratio)_0 = (5.72 \pm 0.07) \times 10^{-6}$ for Erg Chech 002. 
Assuming they both formed from material sampling the canonical $(\alratio)$ ratio, then NWA 11119 formed at $\Delta t_{26} = 3.55 \pm 0.06$ Myr, and Erg Chech 002 at $\Delta t_{26} = 2.29 \pm 0.01$ Myr.
%Regardless of whether they sampled the CAI $\altwosix$ reservoir, 
{\Steve If these values are correct and} if these two achondrites formed from a common reservoir, then clearly NWA 11119 formed a time $\Delta t = \tau_{26} \, \ln \left[ (5.72 \pm 0.07) / (1.69 \pm 0.09) \right]$ $= 1.26 \pm 0.08$ Myr after Erg Chech 002.
However, the only dating information reported by \citet{SrinivasanEtal2018} in the abstract of their paper is that NWA 11119 has an absolute age of $4564.8 \pm 0.3$ Myr. 
Likewise, the only dating information \citet{BarratEtal2021} reported in the abstract of their paper was that Erg Chech 002 has an absolute age $4565.0$ (presumably $\pm 0.3$) Myr.
Based on these headline ages, the age difference is $0.2 \pm 0.4$ Myr, which is clearly incompatible with an age difference of $1.26 \pm 0.08$ Myr. 

Only a {\it very} careful reading of the papers will clear up the discrepancy.
The age $4564.8 \pm 0.3$ Myr for NWA 11119 was derived by \citet{SrinivasanEtal2018} by anchoring to Al-Mg and Pb-Pb ages of D'Orbigny, although at no point did they explicitly state the $(\alratio)_0$ and Pb-Pb age of D'Orbigny they were assuming.
In contrast, the $4565.0$ Myr for Erg Chech 002 was derived by \citet{BarratEtal2021} by anchoring to CAIs, implicitly assuming their Pb-Pb age is 4567.3 Myr, although at no point did they state the assumed Pb-Pb age of CAIs, recognize the collateral assumption that $\altwosix$ would have to be heterogeneous, or even cite the source \citep{ConnellyEtal2012} of these assumptions. 
So many assumptions are introduced when using anchors that the entire meaning of a sample age becomes opaque. 
Chronometry papers are replete with phrases like “if anchored to D’Orbigny, then... but if anchored to CAIs, then...”. 
This is confusing and not even necessary, as the quantity of greatest interest is the time of formation after $t\!\!=\!\!0$. 
Moreover, the use of absolute ages necessarily introduces unneeded uncertainty, because of the uncertainties in even Pb-Pb relative ages, $\pm 0.3$ Myr, or $\pm 0.7$ Myr in the example of NWA 6704 above.
Had the formation times of NWA 11119 and Erg Chech 002 been reported as times after $t\!\!=\!\!0$, the precision in the difference in their formation times would have been $\pm 0.08$, a factor of 4 more precise than could be obtained using the ages reported for them. 

For all these reasons, we strongly advocate reporting all ages, even Pb-Pb ages, as times of formation relative to $t\!\!=\!\!0$, avoiding the use of anchors, and stating all assumptions clearly.
This means reporting the assumed value of $t_{\rm SS}$ for Pb-Pb dating, or the value of $(\alratio)_{\rm SS}$ or even $(\mnratio)_{\rm SS}$ that was assumed.
We strongly advocate avoiding use of anchors to report absolute ages.

\subsection{Improved Chronometry}

Fixing {\Steve $t_{\rm SS} = 4568.42 \pm 0.24$ Myr} allows us to use the Pb-Pb system as a relative chronometer, in concert with the Al-Mg chronometer, improving the precision of dating.  
In {\bf Table~\ref{table:modelages}} we present for various meteoritic samples: our calculated time of formation after $t\!\!=\!\!0$ based on the Al-Mg system, $\Delta t_{26}$; our calculated time time of formation based on the Pb-Pb system and {\Steve $t_{\rm SS} = 4568.42$ Myr}, $\Delta t_{\rm Pb}$; and the weighted mean of these, $\Delta t$.
%We do not report a weighted mean for NWA 2976 or NWA 6704, because the $\Delta t_{26}$ and $\Delta t_{Pb}$ ages do not appear compatible with each other.
To connect to the broader literature, we also list the implied absolute age, based on {\Steve $t_{\rm SS} = 4568.42$} Myr (and the assumption of commonly used U half-lives), but we emphasize again that the relative ages are what matter for chronometry.
{\Steve The uncertainties reflect only the uncertainty in the Al-Mg ages, as is appropriate when taking the differences in model ages.}
%The model age of NWA 1670 appears consistent with the isochron shown in Figure~\ref{fig:nwa1670low}, but not the original one derived by \citet{SchillerEtal2015}.
We also note that in a companion paper (\citealt{DeschEtal2023b}, hereafter Paper II) we update these estimates using information from the Mn-Cr and Hf-W systems, and many more achondrites; these shift the inferred times of formation $\Delta t$ and model ages, by {\Steve about 0.07 Myr}.

%-------------------------------------------
% TABLE 2. Model ages
%$\!\!\!\!\!\!\!\!\!\!\!\!\!\!\!\!\!\!\!\!\!\!\!\!\!\!\!\!\!\!\!\!$
\begin{table}[h!]
\noindent
\begin{minipage}{4.5in}
\footnotesize
\caption{Adopted times of formation $\Delta t_{26}$ and $\Delta t_{\rm Pb}$ (assuming $t_{\rm SS} = 4568.42$ Myr), and weighted mean, $\Delta t$, plus implied Pb-Pb age, of seven bulk achondrites and four chondrules. \vspace{0.1in}
\label{table:modelages}}

\begin{tabular}{l|c|c|c|c}
{Sample} & $\Delta t_{26}$ (Myr) & $\Delta t_{\rm Pb}$ (Myr) & $\Delta t$ (Myr) & Model age \\
\hline
{D'Orbigny} & 
$5.06 \pm 0.10$ & $4.99 \pm 0.21$ & $5.05 \pm 0.09$ & $4563.37 \pm 0.09$ \\
{SAH 99555} &
$5.14 \pm 0.05$ & $4.72 \pm 0.24$ & $5.12 \pm 0.05$ & $4563.30 \pm 0.05$ \\
{NWA 1670} &
$4.64 \pm 0.10$ & $4.21 \pm 0.66$ & $4.63 \pm 0.10$ & $4563.79 \pm 0.10$ \\
{Asuka 881394} & 
$3.82 \pm 0.04$ & $3.47 \pm 0.53$ & $3.82 \pm 0.04$ & $4564.60 \pm 0.04$ \\
{NWA 7325} & 
$5.33 \pm 0.05$ & $4.52 \pm 1.70$ & $5.33 \pm 0.05$ & $4563.09 \pm 0.05$ \\
\hline
{NWA 2796} & 
$5.03 \pm 0.04$ & $5.26 \pm 0.57$  & $5.03 \pm 0.04$ & $4563.39 \pm 0.04$ \\
{NWA 6704} & 
$5.29 \pm 0.13$ & $5.66 \pm 0.26$ & $5.36 \pm 0.11$ & $4563.13 \pm 0.13$ \\
\hline 
{NWA 5697 2-C1} & 
$2.00 \pm 0.21$ & $0.82 \pm 0.75$ & $1.91 \pm 0.20$ & $4566.51 \pm 0.20$ \\
{NWA 5697 3-C5} & 
$1.84 \pm 0.21$ & $2.35 \pm 1.07$  & $1.86 \pm 0.21$ & $4566.56 \pm 0.21$ \\
{NWA 5976 5-C2} & 
$2.07 \pm 0.22$ & $1.18 \pm 0.95$ & $2.03 \pm 0.21$ & $4566.39 \pm 0.21$ \\
{NWA 5697 11-C1} & 
$2.32 \pm 0.34$ & $2.91 \pm 0.74$ & $2.42 \pm 0.31$ & $4566.00 \pm 0.31$ \\
\hline
\end{tabular}

%\footnotesize{}
\end{minipage}
\end{table}
%-------------------------------------------

Having a precise and accurate value for $t_{\rm SS}$ obtained by averaging of multiple samples allows improved chronometry across the board, not just for the samples above with Al-Mg dates, but also for samples with Mn-Cr measurements. 
For example, if a sample has a Pb-Pb age $t_{\rm Pb}$ and a measurement of $(\mnratio)_0$, one could extrapolate backward to find the implied value of $(\mnratio)_{\rm SS}$ at $t\!\!=\!\!0$ in the solar nebula:
\begin{equation}
(\mnratio)_{\rm SS} = (\mnratio)_0 \, \exp \left( + \Delta t_{Pb} / \tau_{53} \right).
\end{equation}
Even samples formed too late to have Al-Mg ages could be used to estimate $(\mnratio)_{\rm SS}$, as \citet{NyquistEtal2009} did using LEW 86010.
However, we advocate using averages of many such samples to derive $(\mnratio)_{\rm SS}$. 
Once it is derived, a single measurement of $(\mnratio)_0$ in a sample, even without a Al-Mg or Pb-Pb date, would suffice to date the formation time of the sample. 
Deriving such quantities is one goal of the companion Paper II \citep{DeschEtal2023b}.

\section{Conclusions}

{\Steve
We have correlated Pb-Pb ages and Al-Mg times of formation of seven achondrites and four chondrules for which both $(\alratio)_0$ and U-corrected Pb-Pb ages $t_{\rm Pb}$ exist. 
We define $t\!\!=\!\!0$ to be the time in the solar nebula when $(\alratio) = 5.23 \times 10^{-5}$ (assuming homogeneity of $\altwosix$). 
The time of formation after $t\!\!=\!\!0$ is then $\Delta t_{26} = \tau_{26} \, \ln \left[ (\alratio)_{\rm SS} / (\alratio)_0 \right]$, where $\tau_{26} = 1.034$ Myr is the mean-life of $\altwosix$. 
The time of formation according to the Pb-Pb system is $\Delta t_{\rm Pb} = t_{\rm SS} - t_{\rm Pb}$, where $t_{\rm SS}$ is the Pb-Pb age of samples that achieved isotopic closure at $t\!\!=\!\!0$, using standard half-lives for ${}^{235}{\rm U}$ and ${}^{238}{\rm U}$. 
The achondrites rapidly cooled and should have achieved simultaneous isotopic closure in both systems. 
In these samples in particular, the times of formation $\Delta t_{26}$ and $\Delta t_{\rm Pb}$ should match if $\altwosix$ was homogeneous. 
These samples therefore test the hypothesis of homogeneity.
Specifically, if a single value of $t_{\rm SS}$ does not reconcile (in a statistical sense) the $\Delta t_{26}$ and $\Delta t_{\rm Pb}$ formation times for the achondrites, this would falsify the homogeneous hypothesis.
}

{\Steve
We find that the hypothesis of $\altwosix$ homogeneity is not falsified. 
The seven achondrites' ages are reconciled by $t_{\rm SS} = 4568.42 \pm 0.24$ Myr.
Thirteen of their 14 formation times are concordant at the $< 2\sigma$ level, and one (the Pb-Pb age of NWA 6704) is discrepant at the $2.4\sigma$ level, consistent with the scatter being due only to measurement errors.
Likewise, the goodness-of-fit parameter for the ensemble is a statistically significant $\chisq = 0.98$ (47\% probability). 
This supports the assumption that $\altwosix$ was homogeneous.
}

{\Steve
Because transient heating events might have reset one chronometer in chondrules but not the other, chondrules do not necessarily test homogeneity.
Nevertheless, they also appear concordant. 
Combining them with the seven achondrites, we find that the 22 ages are reconciled for values of $t_{\rm SS} = 4568.42 \pm 0.24$ Myr. 
Only two of the 22 ages are discordant: the Pb-Pb ages of NWA 6704 (at the $2.3\sigma$ level) and chondrule \textit{2-C1} (at the $2.9\sigma$ level).
Still, $\chisq = 1.36$ for this ensemble, which is statistically significant (12\% probability).
Despite the spread in the chondrule ages, they also appear concordant, further supporting the assumption of homogeneity.
}

{\Steve
There is only a single CAI, \textit{SJ101}, for which a U-corrected Pb-Pb age and an internal Al-Mg isochron both exist.
These ages are quite discrepant. 
Although this has been interpreted in the past as heterogeneity of $\altwosix$, we have demonstrated that the underlying assumption---that the Al-Mg and Pb-Pb systems closed at the same time---is not justified.
If this CAI were subjected to a transient heating event in the solar nebula like those that chondrules experienced, then the Pb-Pb system would be reset, the Al-Mg system in anorthite would be marginally reset, and the Al-Mg system in pyroxene, melilite and spinel would not be modified. 
The Al-Mg system could record the time of the CAI's formation at $t \approx 0$, while the Pb-Pb system would record a time 1-3 Myr later.
The isochron of \textit{SJ101} is exactly consistent with this behavior, making it untenable to argue that the Al-Mg and Pb-Pb systems had to close simultaneously, or that CAIs' Pb-Pb chronometers record $t\!\!=\!\!0$.
}

{\Steve 
As part of our analysis, we reevaluated Pb-Pb ages of all the samples.
We performed the regressions as a check, and averaged the intercepts of the Pb-Pb isochrons and averaged the ${}^{238}{\rm U}/{}^{235}{\rm U}$ ratios to reevaluate the Pb-Pb ages, where appropriate.
We discovered that the isochrons built by 
\citet{ConnellyEtal2008} [SAH 99555], 
\citet{ConnellyEtal2012} [CAIs],
\citet{SchillerEtal2015} [NWA 1670], and \citet{BollardEtal2017} [NWA 5697 chondrules] employed a methodology for selecting points to regress that is prone to confirmation bias, leading to ages that are overly precise.
We have reevaluated the uncertainties in these ages. 
We have also applied the isotopic correction factor $\delta^{238}{\rm U} \approx -0.3\permil$ advocated by \citet{TissotEtal2017}, to account for the fact that pyroxene grains (from which Pb-Pb isochrons are typically built) are isotopically lighter than uranium in the usually-measured whole rock (in which uranium isotopes are usually measured), leading to Pb-Pb ages being overestimated by about 0.19 Myr.
We argue that this correction should not be applied to the achondrites NWA 2976 and NWA 6704 from the CC isotopic reservoir, due to them forming from hydrous magmas.
}

The difficulty of measurements of Pb-Pb ages in individual CAIs, and the controversy over the results, makes clear that statistical approaches like that of \citet{NyquistEtal2009}, \citet{SanbornEtal2019}, {\Steve \citet{PirallaEtal2023}}, and the approach taken here are preferred. 
Averages of many samples are more reliable and lead to greater precision, than use of single anchors. 
While we find {\Steve $t_{\rm SS} = 4568.42 \pm 0.24$ Myr} from the samples above, we encourage further measurements of Al-Mg and Pb-Pb dates for many more systems for which the Al-Mg and Pb-Pb systems are likely to have achieved isotopic closure simultaneously, especially volcanic achondrites.
This will severely test the prediction of homogeneity, but if $\altwosix$ was homogeneous then we expect $t_{\rm SS}$ to be determined even more precisely as more measurements are made.

We advocate a move away from reporting formation times as absolute ages determined by anchoring to samples.
Use of anchors introduces a number of assumptions about the ages, which are rarely stated explicitly in papers. 
At any rate, absolute ages are not needed for models of protoplanetary disks and planet formation, or for most purposes; formation times $\Delta t$ relative to $t\!\!=\!\!0$, or time differences between two events, are demanded.
Relative ages are much more precise than Pb-Pb absolute ages, which have uncertainties of $\pm 9$ Myr due to uncertainties in the ${}^{235}{\rm U}$ half-life. 
The real utility of the Pb-Pb system is as a relative dating system, with precision $\pm 0.3$ Myr.
This requires fixing $t_{\rm SS}$ to define $\Delta t_{\rm Pb} = t_{\rm SS} - t_{\rm Pb}$.
We hope that our finding of {\Steve $t_{\rm SS} = 4568.42 \pm 0.24$ Myr}, will allow precise relative dating of samples using the Pb-Pb system.

This increased precision should open the door to more precise chronometry using other isotopic systems.
In the companion Paper II \citep{DeschEtal2023b}, we {\Steve refine $t_{\rm SS}$ and} show how knowledge of $t_{\rm SS}$ allows us to use statistical approaches to better determine $(\mnratio)_{\rm SS}$ and $(\hfratio)_{\rm SS}$ at $t\!\!=\!\!0$ in the solar nebula, despite the difficulties of measuring these quantities in CAIs directly. 
This allows a measurement of $(\mnratio)_0$ or $(\hfratio)_0$ in a sample to be converted into a time of formation directly, without the need for anchors or absolute ages.

\bigskip
{\bf Acknowledgments}: The authors would like to acknowledge the efforts of cosmochemists from multiple laboratories around the world whose work makes possible the data cited in Table 1 and throughout this paper. Statistical chronometry necessarily distills very difficult and painstaking analytical work into mere numbers to be crunched, but the efforts to obtain those numbers are appreciated. 
We thank Zachary Torrano for useful discussions.
We thank {\EDIT Francois Tissot and two other anonymous reviewers, whose} suggestions greatly improved the quality of our work.
The work herein benefitted from collaborations and/or information exchange within NASA's Nexus for Exoplanetary System Science research coordination network sponsored by NASA's Space Mission Directorate (grant NNX15AD53G, PI Steve Desch).
Emilie Dunham gratefully acknolwedges support from a 51 Pegasi b Fellowship, grant \#2020-1829.

\bigskip

The data in Table 1 and the calculations by which we derived our results are included as an Excel spreadsheet as Research Data.

\bigskip

\bibliographystyle{elsarticle-harv} 
%\bibliography{cas-refs}
\bibliography{AnchorsAway}





% \end{thebibliography}
\end{document}